\documentclass[a4paper,10pt,prd,aps,floatfix,preprintnumbers,twocolumn,nofootinbib,superscriptaddress]{revtex4-1}
\usepackage{amssymb}
\usepackage{amsmath}
\usepackage{graphicx}
\usepackage{hyperref}
\usepackage{orcidlink}
\usepackage{verbatim}
\usepackage{microtype}

\newcommand{\orcidauthorMASON}{0000-0002-1857-1085}
\newcommand{\orcidauthorBENNETT}{0000-0002-1678-6701}
\newcommand{\orcidauthorLUCINI}{0000-0001-8974-8266}
\newcommand{\orcidauthorPIAI}{0000-0002-2251-0111} 
\newcommand{\orcidauthorRINALDI}{0000-0003-4134-809X} 
\newcommand{\orcidauthorVADACCHINO}{0000-0002-5783-5602}
\newcommand{\orcidauthorZIERLER}{0000-0002-8670-4054} 

\newcommand{\beq}{\begin{equation}}
\newcommand{\eeq}{\end{equation}}
\newcommand{\beqs}{\begin{eqnarray}}
\newcommand{\eeqs}{\end{eqnarray}}
\newcommand{\llangle}{\langle \kern-.17em \langle}
\newcommand{\rrangle}{\rangle \kern-.17em \rangle}

\makeatletter
\def\@collaboration@present#1#2#3#4{%
 \par
 \begingroup
  \frontmatter@collaboration@above
  \@affilID@def{}%
  \@tempcnta\z@
  \@author@present{}{\ignorespaces#3\unskip}{#4}%
  \par
 \endgroup
 \set@listcomma@list#1%
}%
\makeatother

\begin{document}
\newcommand\MaxAspectRatioNtFour{12}
\newcommand\MaxAspectRatioNtFive{16}
\newcommand\llrtherm{300}
\newcommand\llrmeas{700}

\author{Ed Bennett\,\orcidlink{\orcidauthorBENNETT}}
\email{e.j.bennett@swansea.ac.uk}
\affiliation{Swansea Academy of Advanced Computing, Swansea University (Bay Campus), Fabian Way, Swansea SA1 8EN, United Kingdom}
\affiliation{Centre for Quantum Fields and Gravity, Faculty  of Science and Engineering, Swansea University, Singleton Park, SA2 8PP, Swansea, United Kingdom}

\author{Biagio Lucini\,\orcidlink{\orcidauthorLUCINI}}
\email{b.lucini@qmul.ac.uk}
\affiliation{Swansea Academy of Advanced Computing, Swansea University (Bay Campus), Fabian Way, Swansea SA1 8EN, United Kingdom}
\affiliation{Department of Mathematics, Faculty of Science and Engineering, Swansea University (Bay Campus), Fabian Way, SA1 8EN Swansea, United Kingdom}
\affiliation{School of Mathematical Sciences, Queen Mary University of London, Mile End Road, London, E1 4NS, UK}

\author{David Mason\,\orcidlink{\orcidauthorMASON}}
\email{d.mason@fz-juelich.de}
\affiliation{Jülich Supercomputing Centre, Forschungszentrum Jülich, D-52425 Jülich, Germany}

\author{Maurizio Piai\,\orcidlink{\orcidauthorPIAI}}
\email{m.piai@swansea.ac.uk}
\affiliation{Centre for Quantum Fields and Gravity, Faculty  of Science and Engineering, Swansea University, Singleton Park, SA2 8PP, Swansea, United Kingdom}
\affiliation{Department of Physics, Faculty  of Science and Engineering, Swansea University, Singleton Park, SA2 8PP, Swansea, United Kingdom}

\author{Enrico Rinaldi\,\orcidlink{\orcidauthorRINALDI}}
\email{erinaldi.work@gmail.com}
\affiliation{RIKEN Center for Interdisciplinary Theoretical and Mathematical Sciences (iTHEMS), RIKEN, 2-1 Hirosawa, Wako, Saitama, 351-0198, Japan}

\author{Davide Vadacchino\,\orcidlink{\orcidauthorVADACCHINO}}
\email{davide.vadacchino@plymouth.ac.uk}
\affiliation{Centre for Mathematical Sciences, University of Plymouth, Plymouth, PL4 8AA, United Kingdom}

\author{Fabian Zierler\,\orcidlink{\orcidauthorZIERLER}}
\email{fabian.zierler@swansea.ac.uk}
\affiliation{Centre for Quantum Fields and Gravity, Faculty  of Science and Engineering, Swansea University, Singleton Park, SA2 8PP, Swansea, United Kingdom}
\affiliation{Department of Physics, Faculty  of Science and Engineering, Swansea University, Singleton Park, SA2 8PP, Swansea, United Kingdom}

\collaboration{(on behalf of the TELOS collaboration) \\ 
\href{https://telos-collaboration.github.io}{ \includegraphics[height=1cm]{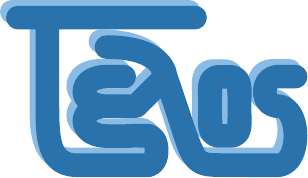} } }
\noaffiliation

\title{Finite-temperature Yang-Mills theories with the density of states method: towards the continuum limit}
\date{\today}

\begin{abstract}
A first-order, confinement/deconfinement phase transition appears in the finite temperature behavior of many non-Abelian gauge theories. These theories play an important role in proposals for completion of the Standard Model of particle physics, hence the phase transition might have occurred in the early stages of evolution of our universe, leaving behind a detectable relic stochastic background of  gravitational waves. Lattice field theory studies implementing the density of states method have the potential to provide  detailed information about the phase transition,  and measure the parameters determining the gravitational-wave power spectrum, by overcoming some of the challenges faced by importance-sampling methods. We assess this potential for a representative choice of Yang-Mills theory with $Sp(4)$ gauge group. We characterize its finite-temperature, first-order phase transition, in the  thermodynamic (infinite volume) limit, for two different choices of number of sites in the compact time direction, hence taking the first steps towards the continuum limit extrapolation. We demonstrate the persistence of non-perturbative phenomena associated to the first-order phase transition:  coexistence of states, metastability, latent heat, surface tension. We find consistency between several different strategies for the extraction of the volume-dependent critical coupling, hence assessing the size of systematic effects. We also determine the minimum choice of ratio between spatial and time extent of the lattice that allows to identify the contribution of the surface tension to the free energy. We observe that this ratio scales non-trivially with the time extent of the lattice, and comment on the implications for  future high-precision numerical studies.
\end{abstract}

\maketitle

\section{Introduction}\label{sec:introduction}

Finite-temperature, first-order phase transitions in gauge theories
are critical to our understanding of particle physics and cosmology. Yet, their characterization on the lattice
requires overcoming algorithmic and technological challenges. It is hence
 the subject of a vast literature, both for $SU(3)$ gauge theories---see for example
 the discussion in Ref.~\cite{borsanyi:2022xml}, the review~\cite{Aarts:2023vsf}, and references therein, in particular
Refs.~\cite{Svetitsky:1982gs,Yaffe:1982qf}, but also Refs.~\cite{Kajantie:1981wh,Celik:1983wz,Kogut:1983mn,
Svetitsky:1983bq,Gottlieb:1985ug,Brown:1988qe,Fukugita:1989yb,Bacilieri:1989ir,
Alves:1990pn,
Boyd:1995zg,Boyd:1996bx,
Borsanyi:2012ve,Shirogane:2016zbf} and 
Refs.~\cite{Saito:2011fs,Ejiri:2019csa,Kiyohara:2021smr,
Fromm:2011qi,Cuteri:2020yke,Dini:2021hug,
Borsanyi:2021yoz}---and for other gauge groups, relevant for new physics~\cite{Lucini:2002ku,Lucini:2003zr,Holland:2003kg,Lucini:2005vg,Pepe:2005sz,Pepe:2006er,Cossu:2007dk,Panero:2009tv,Datta:2010sq,Lucini:2012wq,Bruno:2014rxa,Appelquist:2015yfa,Appelquist:2015zfa,LatticeStrongDynamics:2020jwi,Bruno:2024dha}.
 
Appealing extensions of the Standard Model (SM) of particle physics
postulate the existence of new dark sectors~\cite{Strassler:2006im,Cheung:2007ut,Hambye:2008bq,Feng:2009mn,
Cohen:2010kn,Foot:2014uba,
Bertone:2016nfn}, taking a variety of forms: composite dark matter models ~\cite{DelNobile:2011je,
Hietanen:2013fya,Cline:2016nab,Cacciapaglia:2020kgq,Dondi:2019olm,
Ge:2019voa,Beylin:2019gtw,Yamanaka:2019aeq,Yamanaka:2019yek,Cai:2020njb},  
strongly interacting massive particle (SIMP) models~\cite{Hochberg:2014dra,Hochberg:2014kqa,Hochberg:2015vrg,Bernal:2017mqb,Berlin:2018tvf,
Bernal:2019uqr,Tsai:2020vpi,Kondo:2022lgg,Chu:2024rrv,Maas:2021gbf,Zierler:2021cfa,Kulkarni:2022bvh,Pomper:2024otb}, 
dark dilaton effective field theories~\cite{Cacciapaglia:2023kat,Ferrante:2023bcz,Appelquist:2024koa}.
They address puzzles in the Standard Model of cosmology, such as observational evidence that dark energy and dark matter, that have no SM explanation,  dominate our present universe (see, e.g., the review in Ref.~\cite{Cirelli:2024ssz}). Even explaining the observed baryonic component requires a matter-antimatter asymmetry that admits no dynamical SM origin---see  Refs.~\cite{Laine:1998jb,Morrissey:2012db} for reviews, and references therein. In particular Refs.~\cite{Kajantie:1996mn,Karsch:1996yh,Gurtler:1997hr,
Rummukainen:1998as,Csikor:1998eu,Aoki:1999fi,
DOnofrio:2015gop,Gould:2022ran} show the failure of the Sakharov's out-of-equilibrium condition~\cite{Sakharov:1967dj} for electroweak baryogenesis.

If the dark sector gauge theory undergoes, in the early universe, a strong  first-order phase transition, it leads to the emission of gravitational waves~\cite{Witten:1984rs,Kamionkowski:1993fg,Allen:1996vm,Schwaller:2015tja,Croon:2018erz,Christensen:2018iqi}, potentially detectable by
 future experimental programmes~\cite{Seto:2001qf,
 Kawamura:2006up,Crowder:2005nr,Corbin:2005ny,Harry:2006fi,
 Hild:2010id,Yagi:2011wg,Sathyaprakash:2012jk,Thrane:2013oya,
 Caprini:2015zlo,
 LISA:2017pwj,
 LIGOScientific:2016wof,Isoyama:2018rjb,Baker:2019nia,
 Brdar:2018num,Reitze:2019iox,Caprini:2019egz,
 Maggiore:2019uih}---see also Ref.~\cite{afzal2023nanograv}.
 Appraising the viability of experimental searches
necessitates a precise characterization
of the physics in proximity of the transition, including a measurement of 
the parameters that control the gravitational wave (GW) power spectrum. Of particular interest are
$\alpha$, which is a measure of the strength of the transition, and $\beta/H_{\ast}$~\cite{Caprini:2019egz}, where $\beta$ corresponds to the duration of the phase transition and $H_{\ast}$ is the Hubble parameter at the transition. The latent heat can be related to $\alpha$ and the bubble nucleation rate (and bubble surface tension) can be related to $\beta/H_{\ast}$.\footnote{Calculating the  nucleation rate is itself a challenging task~\cite{Kalikmanov2013}.}

The main challenge for the direct lattice numerical calculation of these quantities
(that act as input values in numerical packages such as PTPlot~\cite{Caprini:2019egz})
is that first-order phase transitions are accompanied by phase coexistence and metastability. These phenomena compromise ergodicity and detailed balance requirements of traditional importance-sampling techniques, based on Markov chain Monte Carlo update algorithms, and lead to critical slowing down~\cite{Berg:1991cf}.
Indirect routes
can be pursued to estimate the power spectrum 
in strongly coupled theories~\cite{Huang:2020crf,Halverson:2020xpg,Kang:2021epo,Reichert:2021cvs,Reichert:2022naa,Pasechnik:2023hwv},
using effective
 Polyakov-loop theory~\cite{ Pisarski:2000eq,
Pisarski:2001pe,Pisarski:2002ji,Sannino:2002wb,
Ratti:2005jh,Fukushima:2013rx,Fukushima:2017csk,Lo:2013hla,Hansen:2019lnf}
and matrix models~\cite{Meisinger:2001cq,Dumitru:2010mj,
Dumitru:2012fw, Kondo:2015noa,Pisarski:2016ixt,Nishimura:2017crr,Guo:2018scp,
KorthalsAltes:2020ryu,Hidaka:2020vna}.
Most recently, gauge-gravity duality techniques~\cite{Maldacena:1997re,Gubser:1998bc,Witten:1998qj,Aharony:1999ti},
adapted to  capture confinement and chiral symmetry breaking~\cite{Witten:1998zw,Klebanov:2000hb,Maldacena:2000yy,Chamseddine:1997nm,Babington:2003vm,Butti:2004pk,Brower:2000rp,Karch:2002sh,Kruczenski:2003be,Sakai:2004cn,Sakai:2005yt}, have also been used for the treatment of first-order phase transitions in strongly coupled field theories---see, e.g., Refs.~\cite{Bigazzi:2020phm,Ares:2020lbt,Bea:2021zsu, Bigazzi:2021ucw,Henriksson:2021zei,Ares:2021ntv,Ares:2021nap,Morgante:2022zvc,Bea:2021zol, Bea:2022mfb,Bea:2024xgv,Bea:2024bxu,Bea:2024bls}.

This paper adopts a new, alternative strategy to gain direct access to the proximity of first-order phase transitions, that takes inspiration from flat histogram~\cite{Berg:1991cf} and density of states~\cite{Wang:2000fzi} methods. The Logarithmic Linear Relaxation (LLR) algorithm~\cite{Langfeld:2012ah,Langfeld:2013xbf,Langfeld:2015fua,Cossu:2021bgn}---see also Ref.~\cite{DElia:2024qxa}---is precisely described in the body of the paper. The underlying idea is to scan the space of configurations with a density function inspired by the microcanonical (rather than canonical) ensemble in statistical mechanics, and treat on equal footing stable, metastable, as well as unstable configurations of the lattice system, overcoming the difficulties connected with importance sampling methods. A number of  increasingly sophisticated analyses have already been performed, pertaining to Abelian gauge theories~\cite{Langfeld:2015fua}, and to $SU(3)$ theories at zero~\cite{Cossu:2021bgn} and finite temperature~\cite{Mason:2022trc,Mason:2022aka,Lucini:2023irm}. Finite-temperature studies exist also for $Sp(4)$~\cite{Bennett:2024bhy}, $SU(4)$~\cite{Springer:2021liy,Springer:2022qos} and $SU(N_c)$~\cite{Springer:2023wok,Springer:2023hcc}, all of which yield encouraging preliminary results.

We consider a non-Abelian gauge theory---the $Sp(4)$ Yang–Mills theory in four dimensions---characterize its finite-temperature deconfinement transition using the LLR algorithm, and perform the first scaling test to approach the continuum and infinite-volume limits.
To this purpose, we discretize the theory on hypercubic  lattices with varying number of sites on the temporal, $N_t=4$ and $N_t=5$, and spatial, $N_s$, directions, measure a set of observables, extrapolate them to the physical limits, and assess the magnitude of methodological systematic uncertainties.
The free energy density of the theory displays its multivalued nature in a region of parameter space  in proximity of the first-order phase transition.
We estimate the critical couplings at finite volume, $\beta_{CV}$, (in the following referred to as critical couplings, for brevity) extracted in several complementary ways, with finite spatial volume of the lattice.  We also measure the specific heat and surface tension.
We study in detail how all the lattice observables converge to continuum field theory quantities in the limits of $N_s\rightarrow +\infty$ (the infinite-volume, or thermodynamic, limit) and $N_t\rightarrow +\infty$.

The TELOS collaboration has been carrying out a systematic programme of lattice studies 
of the $Sp(2N)$ gauge theories, both in the pure gauge case~\cite{Bennett:2017kga,Bennett:2020hqd,
Bennett:2020qtj, Bennett:2022gdz, Bennett:2022ftz},
as well as in the presence of fermion matter fields~\cite{Bennett:2017kga,Lee:2018ztv,
  Bennett:2019jzz, Bennett:2019cxd, 
  Bennett:2022yfa, 
  Bennett:2023gbe, Bennett:2023mhh, Bennett:2023qwx, Bennett:2024cqv,
  Bennett:2024wda, Bennett:2024tex,  Bennett:2025amc, Bennett:2025domainwall, Bennett2025scattering,
  Bennett:2025singlets}, reshaping our understanding of  these theories---see also  
  the review in Ref.~\cite{Bennett:2023wjw}, as well as Refs.~\cite{Hong:2017suj,Kulkarni:2022bvh,Bennett:2023rsl, Dengler:2024maq}.
The $Sp(4)$ Yang-Mills theory  is the most accessible and best understood
  one in this class, and hence a natural target of the present study, though our results  are expected to be relevant also for the study of other gauge groups (see Ref.~\cite{Lucini:2023irm} for $SU(3)$).
The study presented in this paper uses some of the 
  technology TELOS developed and made publicly available, such as the implementation within 
the HiRep code~\cite{DelDebbio:2008zf,HiRepSUN} of the  adaptations needed for the study of  $Sp(N_c=2N)$ theories~\cite{Bennett:2017kga,HiRepSpN},  and the implementation of the heat bath and domain decomposition 
initially implemented for $SU(N_c)$ gauge groups~\cite{Lucini:2023irm,mason_HiRep_LLR_v1.0.0},  extended to symplectic gauge groups~\cite{Bennett:2024bhy,mason_HiRep_LLR_v1.1.0}.

The paper is organized as follows.
In Sect.~\ref{sec:dos_llr}, we outline the methodology based on the density of states. We describe pedagogically the relevant aspects of the density of states methods, and the implementation of the LLR algorithm in the (continuum) field theory context of relevance. In Sect.~\ref{sec:llr_lattice}, we discuss the numerical lattice field theory of interest and the formulation of the LLR algorithm on a finite lattice. We report our new results in Sec.~\ref{sec:results}. This section contains also critical discussions, which include direct comparisons to the existing, published data, obtained with smaller time and space extents of the lattice. We conclude in Sec.~\ref{sec:conclusion}, by summarizing the main results of the study, and outlining future avenues for investigation. We collect some technical aspects of the numerical study in the Appendix. 

\section{Field theory formulation of the LLR algorithm} \label{sec:dos_llr}

In this section, we introduce the density of states methodology for non-Abelian gauge theories,
and its implementation in the LLR algorithm. The presentation is self-contained and pedagogical, and
we refer to the literature for additional details~\cite{Langfeld:2012ah,Langfeld:2013xbf,Langfeld:2015fua,Cossu:2021bgn,DElia:2024qxa}.
We also discuss our implementation of the replica exchange, to enforce ergodicity, and comment on our estimation of errors.

\subsection{Density of states} \label{ssec:dos}

We write the partition function of the Yang-Mills gauge theory with group $Sp(2N)$, treated in isolation from
any other fields (including SM ones) as follows:
\begin{align}\label{eq:continuum_partiotion_function}
    Z(\beta) \equiv \int  \left[D A\right] e^{- \beta S[A]}\,,
\end{align}
where $\beta\equiv 4N/g_0^2$, with $g_0$  the bare gauge coupling, and $ \left[D A\right]$ the (Haar) measure.  The Euclidean action, $S[A]$, is expressed
 as a function of the gauge fields, $A=\sum_B A^B_{\mu}T^B$, where $T^B$ are the generators of the group, while
\begin{align}
\label{eq:continuum_action}
    S[A] \equiv \int {\rm d}^4 x \,{\rm Tr} \left[ \frac{1}{2} F_{\mu \nu}(x) F_{\mu \nu}(x) \right]\,, 
\end{align}
and the field-strength tensor is
\begin{align}
    F_{\mu \nu}(x) \equiv \partial_\mu A_\nu(x) - \partial_\nu A_\mu(x) + i [A_\mu(x)\,,\,A_\nu(x)]\,.
\end{align}
Notice how we made explicit the dependence on $\beta$ in the exponential weight, rather than in the action.

The expectation value of a generic operator, $O[A]$, that depends on the gauge fields, is
\begin{align}\label{eq:op_expectation_value}
    \langle O \rangle_{\beta}  \equiv \frac{1}{Z(\beta)} \int  \left[D A\right] O[A] e^{- \beta S[A]}\,.
\end{align}
This expression implicitly defines a probability density  in the space of the fields, $ {\rm d}P_\beta[A]$,  taking the form
\begin{align}
    {\rm d}P_\beta(A) = \frac{1}{Z(\beta)}  \left[D A\right] e^{-\beta S[A]}\,.
\end{align}
In the presence of a first-order transition, this probability density  may show multiple, inequivalent stationary points, corresponding to coexisting phases. On the lattice, importance-sampling (Markov chain) algorithms have exponentially suppressed transition rates in Monte-Carlo time between regions near different local maxima of the probability distribution. The resulting freezing of the updates near one of the local maxima introduces the aforementioned violations of ergodicity and detailed balance (hysteresis). Ultimately, the resulting critical slowing down~\cite{Berg:1991cf} makes it unfeasible to overcome the problem with realistic amounts of computational resources.

The  approach based upon the density of states, $\rho(E)$, is designed to overcome these challenges. 
One constrains the action, $S[A]$, to match the energy, $S[A]=E$:
\begin{align} \label{eq:dos}
    \rho(E) \equiv \int  \left[D A\right] \delta( S[A] - E)\,,
\end{align}
and then rewrites the ensemble average of any
 operator that is only a function of the action $O(E=S[A])$ as\footnote{This approach can be generalized to other observables, such as the Polyakov loop, in which case additional information may be required---see, e.g., the discussion in Ref.~\cite{Langfeld:2015fua}.}
\begin{align}\label{eq:op_dos_expectation_value}
    \langle O \rangle_{\beta}  = \frac{1}{Z(\beta)} \int {\rm d}E \rho(E) O(E) e^{- \beta E}\,.
\end{align}
The calculation of $\langle O \rangle_{\beta} $ for a value of $\beta$ therefore requires determining the density of states, $\rho(E)$, at all values of the energy $S[A]=E$. As in the process we explore all values of $E$ individually, there is no concern about transitions
between inequivalent configurations (phases), that have different energy, $E$.
The probability distribution density for a given energy is then written as
\begin{align}\label{eq:probability_density}
    P_\beta(E) = \frac{1}{Z(\beta)} \rho(E) e^{-\beta E}\,.
\end{align}
This method can yield a precise determination of the critical coupling, since $\beta$ is no longer a parameter entering the Monte Carlo calculations, but rather can be tuned freely 
once $\rho(E)$ has been determined numerically. The non-trivial algorithm leading to 
such determination is the subject of the next subsection.

\subsection{Linear logarithmic relaxation (LLR)} \label{ssec:llr}

When computing ensemble averages with Eq.~(\ref{eq:op_dos_expectation_value}), one finds that they are dominated by integrations over finite energy intervals. We use this empirical observation to
guide our heuristic choice of  the range $\left[E_{\min},E_{\max}\right]$, which includes all such intervals, 
but requires deploying only limited amounts of computational resources,
as we neglect exponentially suppressed contributions from other 
energies. A caveat to this approach is that it compromises the implementation of the 
third law of thermodynamics, which would require the continuous mapping of  $\rho(E)$ to reach the low energy regime close to zero temperature. The drawback is the appearance of an arbitrary  constant, to which we return shortly.

We cover the energy range,  $\left[E_{\min},E_{\max}\right]$, with  intervals 
of finite width, $\Delta_E/2$,\footnote{We follow the notation from Ref.~\cite{Bennett:2024bhy}.}
centered around regularly spaced energies, $E_0^{(n)}$, for $n=1,\, \dots,\, N$.
We characterize the density of states in terms of its logarithmic derivative,  $a(E)$, defined as
\beqs\label{eq:a_of_E}
    a(E) \equiv \frac{\rm d}{{\rm d} E} \log \rho (E)\,,
\eeqs
which we determine by modelling it with the approximation $\log \tilde{\rho}(E)\simeq \log \rho(E)$,  
consisting of a piecewise-linear function
defined so that
\beqs\label{eq:llr_approx}
    \log \rho(E) &\simeq &    \log \tilde{\rho}(E) \,\equiv\,  a^{(n)} \left(E_0^{(n)} - E \right) + c^{(n)}\,,\\
    &{\rm for}& \nonumber
    E \in \left[ E_0^{(n)} - \frac{\Delta_E}{4}, E_0^{(n)} + \frac{\Delta_E}{4} \right]\,.
\eeqs
The  determination of $\tilde{\rho}$ is then equivalent to the reconstruction of all $a^{(n)}$, and $c^{(0)}$. 
The coefficients $c^{(n)}$, with $n>0$, are determined by requiring continuity. The first coefficient, $c^{(0)}$, is left undetermined by this approach, but  as anticipated only results in an overall normalization of the partition function, $Z(\beta)$, that drops out of ensemble averages such as those in Eq.~(\ref{eq:op_dos_expectation_value}).

In order to illustrate how we determine the numerical values of $a^{(n)}$, we introduce the restricted expectation value, for which we use the double-bracket notation:
\begin{widetext}
\beqs\label{eq:double_bracket}
    \llangle f \rrangle_n (\hat a) &\equiv& 
    \frac{1}{{\cal N}_n(\hat{a}) }\!\int\! D[A] f[A] e^{- \hat a \left(S[A]\!-\!E_0^{(n)}\right)} W(S[A]\!-\!E_0^{(n)}\!,\delta) \,,\\
  {\rm where}~~~~~{\cal N}_n(\hat{a}) &\equiv& \int\! D[A] e^{- \hat a \left(S[A]-E_0^{(n)}\right)} W(S[A]-E_0^{(n)},\delta)\,,\\
   {\rm and}~~~~W(x,\delta) &\equiv& 
    \begin{cases}
        1\,, &\text{if} \quad -\delta/2 < x < \delta/2 \\
        0\,, &\text{otherwise}
    \end{cases}\,,
\eeqs
\end{widetext}
for any function, $f(A)$,  of the fields, $A$.\footnote{The double bracket in the notation reflects the fact that this object closely resembles the simultaneous  application of microcanonical and  canonical ensembles, as it amounts to restricting the integration to a narrow interval around a given value of the energy, $E_0^{(n)}$, but also to multiplying the integrand by a weight function depending exponentially on $S[A]$---an overall factor of $e^{\hat{a}E_0^{(n)}}$ drops from the ratio with the normalization, ${\cal N}_n(\hat{a})$.}
Here, $\hat a$ is a generic parameter, while $W(x,\delta)$  restricts the energy to an interval centered around $x$ with  
fixed width, $\delta$.

As the double-bracket expectation value, $\llangle f \rrangle_n(\hat a)$, in Eq.~(\ref{eq:double_bracket}) can be explicitly rewritten in the form of Eq.~\eqref{eq:op_expectation_value}, it can be calculated using standard Monte-Carlo techniques. But because the energy is restricted to an  interval, this calculation (for sufficiently small choices of interval, $\delta$) is not affected by the aforementioned challenges, due to possible metastability and phase coexistence.
Furthermore, it can be shown that~\cite{Langfeld:2012ah}
\begin{align}\label{eq:double_bracket_llr}
    \llangle S[A] - E_0^{(n)} \rrangle_n (\hat a = a^{(n)}) = 0\,.  
\end{align}
This observation allows to translate the problem of determining $a^{(n)}$ (and the density of states) into the algebraic 
solution of Eq.~(\ref{eq:double_bracket_llr}), for each 
interval labeled by $n$. 
We do so iteratively, with  a combination of the Newton-Raphson (NR) and Robbins-Monro (RM) updates  as defined below~\cite{RobbinsMonro1951}---see also Tab.~\ref{tab:llr_runs} for the exact number of the respective number of iterations.
 After estimating the derivative of  $\llangle S[A] - E_0^{(n)} \rrangle_n$ \cite{Langfeld:2012ah,Langfeld:2015fua}, we arrive at the iterative equation
\begin{align}\label{eq:llr_update}
    a^{(n)}_{k+1} = a^{(n)}_{k} - \alpha_{k+1} \frac{12}{\delta^2}\llangle S[A] - E_0^{(n)} \rrangle_n ( a^{(n)}_{k} ),
\end{align}
where $\alpha_k = 1$ for the standard Newton-Raphson method, and $\alpha_k = 1/k$ for the Robbins-Monro updates. The Robbins-Monro algorithm guarantees that the iteration prescription converges for stochastic estimations, which we use when calculating the double-bracket expectation values, as $\sum_k \alpha_k \to \infty$ while $\sum_k \alpha_k^2$ remains finite~\cite{RobbinsMonro1951}.

\subsection{Ergodicity and replica exchanges}

A potential ergodicity flaw of the algorithm ensues from the fact that 
configurations with an energy in the same interval might only be connected via local updates by allowing intermediate energies outside the interval
which is not possible within the framework defined by  Eq.~\eqref{eq:double_bracket}.
Following Ref.~\cite{Langfeld:2015fua},
we apply the \textit{replica exchange method}~\cite{Swendsen:1986vqb,Vogel_2018} to overcome this impasse. The number of replicas equals that  of energy intervals used in the algorithm, $N_{\rm rep}=N$. Furthermore, we choose the energy intervals to overlap with the neighboring ones and simulate all intervals in parallel, 
by fixing the central energies to be spaced by half the interval width~\cite{Lucini:2023irm,Bennett:2024bhy}, so that
\begin{align}
    E_0^{(n+1)} - E_0^{(n)} = \Delta_E/2\,,
\end{align}
and setting $\delta=\Delta_E$.

When  two Markov chains of neighbor intervals are in the overlap region, we swap them with a probability of
\begin{align}
    P_{\rm swap} = \min \left( 1 , e^{\left(S[A^{(n)}] - S[A^{(n+1)}]\right)\left(a^{(n)} - a^{(n+1)}\right)} \right).
\end{align}
We further modify Eq.~\eqref{eq:double_bracket} for the  first and last energy intervals, $n=1$ and $n=N=N_{\rm rep}$, by allowing the Monte Carlo to probe energies outside the range $[E_{\min},E_{\max}]$,  with probability distribution proportional to the Boltzmann weight $\exp \left(-\beta S[A]\right)$, as in typical  importance sampling calculations. This allows the Markov chains to probe all possible energies and thus ensure ergodicity.

\subsection{Error estimation}

In our numerical calculations, we terminate the iteration  in Eq.~\eqref{eq:llr_update} after a fixed number of updating steps. 
We initialise the system by performing a fixed number of Newton-Raphson updates, $n_{\rm NR}$. 
We follow this by a fixed number of Robbins-Monro updates,
 $n_{\rm RM}$
We estimate the error by repeating the same iteration prescription multiple times, while starting from a different random configuration. Error estimates are then obtained using a jackknife analysis over the final result of the repeated calculations, $N_{\rm repeats}$.

\section{Lattice formulation} \label{sec:llr_lattice}
The ensemble averages entering Eq.~(\ref{eq:double_bracket_llr}), and hence the reconstruction of the density of states, are computed by discretising the theory on a hypercubic lattice with finite  spacing, $a$, and space-time  volume, $\tilde{V} = (N_t a)\times(N_s a)^3$. We impose periodic boundary conditions in all directions, and interpret the  temporal extent of the lattice,
 characterized by $N_t<N_s$, in terms of the temperature of the system at equilibrium.

We adopt the Wilson gauge action, expressed in terms of the plaquette, $U_{\mu\nu}(x)$, itself a function of a (gauge) link configuration as
\begin{align}
  S[U] \equiv  \frac{1}{N_c}\sum_{x}\sum_{\mu < \nu} {\rm Re}\, {\rm Tr} \left[1 - U_{\mu\nu}(x) \right]\,.
\end{align}
It is notationally convenient to express the action, $S[U]$, in terms of the average value of the plaquette 
\beqs
u_p &\equiv& \frac{a^4}{6\tilde{V}}  \frac{1}{N_c}\sum_{x}\sum_{\mu < \nu} {\rm Re}\, {\rm Tr} \left[U_{\mu\nu}(x) \right]\,,
\eeqs
where $N_c=4$, for $Sp(4)$, 
and use the equivalent expression
\begin{align}
\label{Eq:up}
    S[U] = \frac{6\tilde{V}}{a^4} \left( 1 - u_p[U] \right).
\end{align}
Our measurements are characterized by the extent of the lattice, the range of energies considered---equivalently, 
the average plaquette, $u_p$---the number of intervals/replica, $N=N_{\rm rep}$, and repeats. 
We report the lattice sizes and parameters used in this work in Tab.~\ref{tab:llr_runs}, in which we indicate explicitly also the number 
of  NR and RM iterations, for each lattice. As anticipated, the HiRep code~\cite{DelDebbio:2008zf,HiRepSUN}, supplemented with the symplectic adaptations~\cite{Bennett:2017kga,HiRepSpN}, has been used, in particular for the domain decomposition of the heat bath 
algorithm~\cite{Bennett:2024bhy,mason_HiRep_LLR_v1.1.0}. We decompose each replica into four domains, along one of the spatial dimensions, as in Ref.~\cite{Bennett:2024bhy}. Every double-bracket expectation value is based on a restricted Monte Carlo Markov chain. We first perform \llrtherm~updates to thermalize the energy-restricted Markov chain. We then measure the operator of interest on 
\llrmeas~thermalized configurations, generated using the restricted heat bath method. 

\begin{table}
    \centering
    \caption{Characterization of the lattice studies used for this work. We report the lattice size, $N_t \times N_s^3$, the energy range, given in terms of the average plaquette, $\left[ u_p^{\min}, u_p^{\max} \right]$, as well as the number of replicas/intervals, $N_{\rm rep}=N$, and the number of repeats, $N_{\rm repeats}$. We further report the total number of Newton-Raphson and Robbins-Monro iteration steps, $N_{\rm NR}$ and $N_{\rm RM}$, respectively. We also include the ensembles from Ref.~\cite{Bennett:2024bhy}, for $N_t=4$, which we use to recalculate all quantities considered in this paper to facilitate comparison with the new data at $N_t=5$.}
\begin{tabular}{|c|c|c|c|c|c|c|c|} \hline
$N_t$ & $N_s$ & $u_{p}^{\rm min}$ & $u_{p}^{\rm max}$ & $N_{\rm rep}$ & $N_{\rm repeats}$ & $n_{\rm NR}$ & $n_{\rm RM}$ \\ \hline \hline 
5 & 48 & 0.588 & 0.592 & 48 & 25 & 10 & 60 \\
5 & 48 & 0.588 & 0.592 & 96 & 25 & 10 & 50 \\
5 & 56 & 0.588 & 0.592 & 128 & 25 & 10 & 50 \\
5 & 56 & 0.588 & 0.592 & 48 & 25 & 10 & 50 \\
5 & 56 & 0.588 & 0.592 & 96 & 25 & 10 & 50 \\
5 & 64 & 0.588 & 0.592 & 95 & 20 & 7 & 50 \\
5 & 72 & 0.588 & 0.592 & 95 & 20 & 11 & 50 \\
5 & 80 & 0.588 & 0.59 & 64 & 20 & 15 & 30 \\
\hline \hline 
4 & 20 & 0.565 & 0.58 & 64 & 20 & 10 & 300 \\
4 & 24 & 0.565 & 0.58 & 64 & 20 & 10 & 300 \\
4 & 28 & 0.565 & 0.58 & 64 & 20 & 10 & 200 \\
4 & 40 & 0.568 & 0.576 & 128 & 25 & 10 & 100 \\
4 & 48 & 0.568 & 0.576 & 128 & 26 & 10 & 100 \\
\hline \hline
\end{tabular}

    \label{tab:llr_runs}
\end{table}

\subsection{Thermodynamics}

Borrowing notation from Ref.~\cite{Bennett:2024bhy}, we can make use of the following identifications, that bring the system back into a form
 familiar from statistical mechanics. We define the entropy as 
\beqs
    s &\equiv & \log\left(\rho\right)\,,
\eeqs
noting that this definition is valid up to an additive constant---see the earlier  discussion about the constant $c^{(0)}$. 
The energy, $E=S$, is identified with the internal energy, and hence the temperature, $t$, is given by
\beqs
  t &\equiv & \frac{\partial E}{\partial s} \,=\, \frac{1}{a^{(n)}}\,.
\eeqs
A Legendre transform yields the free energy, $F$,  as
\begin{align}\label{eq:free_energy}
    F &= E - t s\,.
\end{align}
A second additive constant appears in this definition, and, following Ref.~\cite{Bennett:2024bhy}, we conventionally 
set it so that $F$ vanishes at criticality.

Having reconstructed the density of states, and by making use of Eq.~(\ref{eq:probability_density}), 
we can provide a first measurement of the critical coupling, $\beta_{CV}(P)$, for a given value of $N_t$ and $N_s$, by dialing $\beta$ so that the probability distribution of the partition function, $P_\beta(E)$---equivalently the probability distribution of the average plaquette, $P_\beta(u_p)$, obtained by inverting Eq.~(\ref{Eq:up})---displays a double Gaussian shape, with two peaks of equal height at different values of $u_p$.\footnote{ 
We could as well require the appearance of two peaks with height ratio fixed to a difference conventional factor. As long as the thermodynamic limit leads to two $\delta$-functions, the same critical coupling will be recovered---see Appendix~\ref{sec:peak_ratios}.}

\begin{figure}[t]
    \centering
    \includegraphics[scale=0.577]{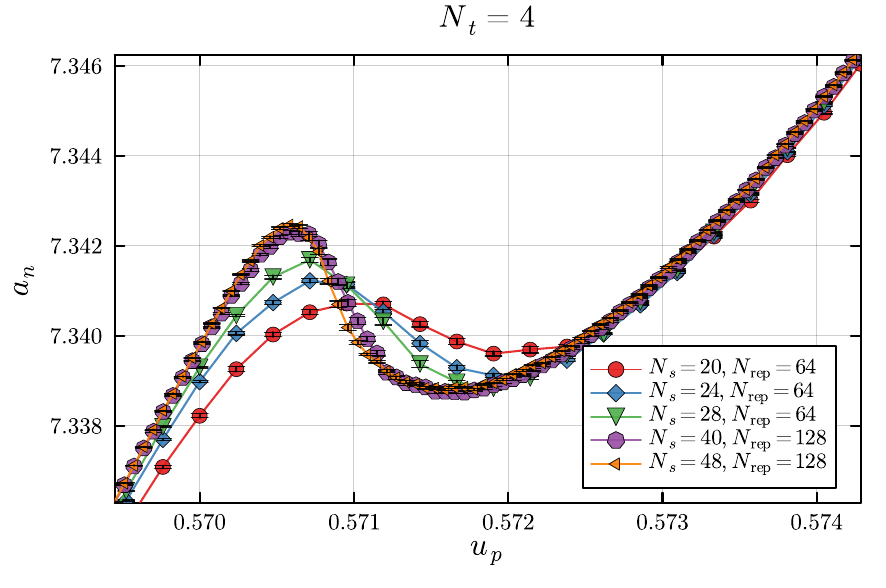}
    \includegraphics[scale=0.577]{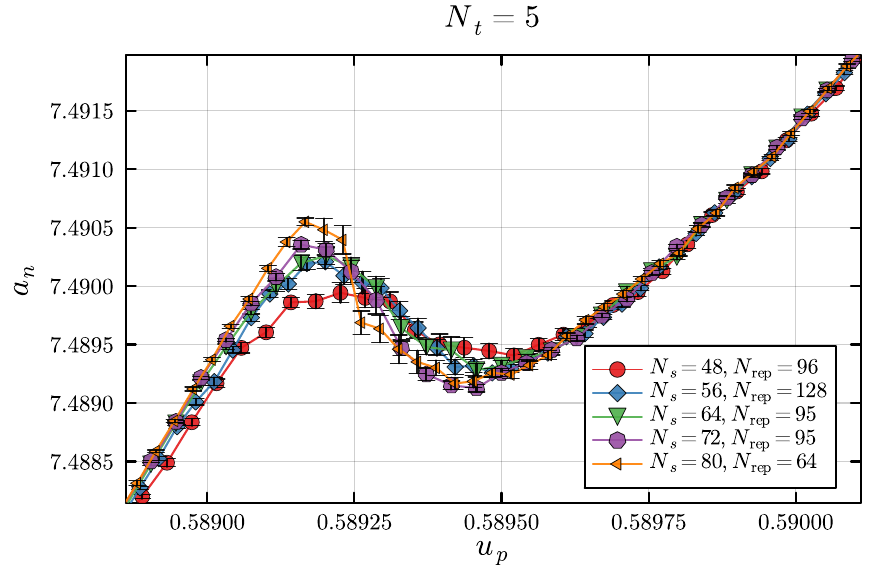}
    \caption{Results for the coefficients, $a^{(n)}$, entering the  piecewise-linear approximation of the logarithm of the density of states, as a function of the average plaquette, $u_p$, computed in the middle of each finite energy interval, $\Delta_E/2$. In both cases with $N_t=4$ (top panel) and $N_t=5$ (bottom), we show results for all available volumes. For volumes for which we performed the calculations with more than one choice of the interval size, we display only the results with the smallest $\Delta_E$.}
    \label{fig:an_volumes}
\end{figure}

Alternative definitions of critical coupling rely on other observables. For example, we consider both the specific heat, $C_V$, and the Binder cumulant, $B_V$. Both can be expressed in terms of moments of the average plaquette,  given, respectively, by
\begin{align}
    \label{eq:specific_heat}
    C_V(\beta) &\equiv \frac{6V}{a^4} \left[ \langle u_p^2 \rangle_\beta - \langle u_p \rangle_\beta^2 \right], \\ 
    \label{eq:binder_cumulant}
    B_V(\beta) &\equiv 1 - \frac{\langle u_p^4 \rangle_\beta}{3\langle u_p^2 \rangle_\beta^2}.
\end{align}
At the critical coupling, the specific heat displays a maximum, whereas the Binder cumulant exhibits a minimum. 
The resulting definition of $\beta_{CV}(C_V)$ and $\beta_{CV}(B_V)$ converge to the same physical temperature in the thermodynamic and continuum limit as $\beta_{CV}(P)$, hence the study of the discrepancy between them provides a useful indicator of the methodological systematics appearing in our calculations.

Assuming that the aspect ratio, $N_s/N_t$, is sufficiently large, we can extract the surface tension, $\sigma_{cd}$, of the bubble walls separating different phases at the confinement/deconfinement transition. Following Ref.~\cite{Lucini:2005vg, Bennett:2025neg}, the interface tension is related to the minimum and maximum of the probability distribution obtained by tuning $\beta$ to the critical value defined by the appearance of peaks with equal heights. One expects that the probabilities computed at the peaks and at the minimum in the valley between them scale as follows:
\begin{align}
    \frac{P_{\min}}{P_{\max}} \propto \sqrt{N_s} \exp \left( -2 \left( \frac{N_s}{N_t} \right)^2 \frac{\sigma_{cd}}{T_c^3} \right)\,,
\end{align}
where $T_c$ is the critical temperature,
In our numerical data, we study the quantity~\cite{Lucini:2005vg}:
\begin{align}\label{eq:surface_tension}
    \tilde{I} = -\frac{1}{2} \left( \frac{N_t}{N_s} \right)^2 \log \left( \frac{P_{\min}}{P_{\max}} \right) + \frac{1}{4} \left( \frac{N_t}{N_s} \right)^2 \log(N_s)\,.
\end{align}
In the thermodynamic limit, $\lim_{N_s/N_t \to \infty} \tilde{I}=\sigma_{cd}/T_c^3$.

\begin{figure}[t]
    \centering
    \includegraphics[scale=0.577]{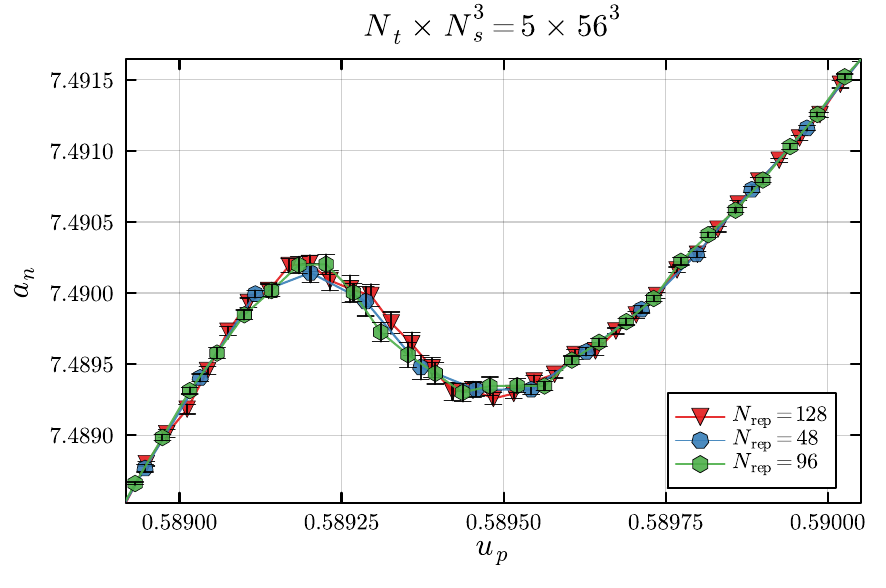}
    \caption{Results for the coefficients, $a^{(n)}$, entering the  piecewise-linear approximation of the logarithm of the density of states, as a function of the average plaquette, $u_p$, computed in the middle of each finite energy interval, $\Delta_E$.
    Having fixed the other parameters, for the representative choice of lattice with extent $N_t\times N_s^3= 5\times 56^3$, we compare three different choices of  number of replicas (and intervals), $N_{\rm rep}$, and hence different energy interval sizes, $\Delta_E$. We observe no statistically significant difference between the sequences of $a^{(n)}$ obtained with these alternative choices of  interval sizes.}
    \label{fig:interval_comparison}
\end{figure}

\section{Numerical results}
\label{sec:results}

We show, in Fig.~\ref{fig:an_volumes}, our results for the values of the coefficients, $a^{(n)}$, entering the approximation of the density of states, as a function of the central value of the plaquette, $u_p$, for lattices with temporal extent $N_t=4$ and $N_t=5$. The former are exhibited for comparison purposes~\cite{Bennett:2024bhy}. The sequence of values of $a^{(n)}$ approximates a  multivalued (non-invertible) function of $u_p$, over a finite range of $u_p$. The difference between the peak and valley values of $a^{(n)}$ inside this non-invertibility range  is less pronounced in the new, $N_t=5$, measurements---roughly by a factor of three. Furthermore, the presence of a range over which $a^{(n)}$ is not invertible as a function of $u_p$ emerges at larger values of $u_p$, and is restricted to a narrower range. It further emerges only at much larger aspect ratios, up to $N_s/N_t=\MaxAspectRatioNtFive$, to become clearly discernible. In existing data on $N_t=4$ only aspect ratios up to $N_s/N_t=\MaxAspectRatioNtFour$ are available, yet the signal of metastability is clearly visible. To the best of our knowledge, the values of aspect ratios used for this publication are the largest ones deployed in published studies of this type.

\begin{figure}[t]
    \centering
    \includegraphics[scale=0.577]{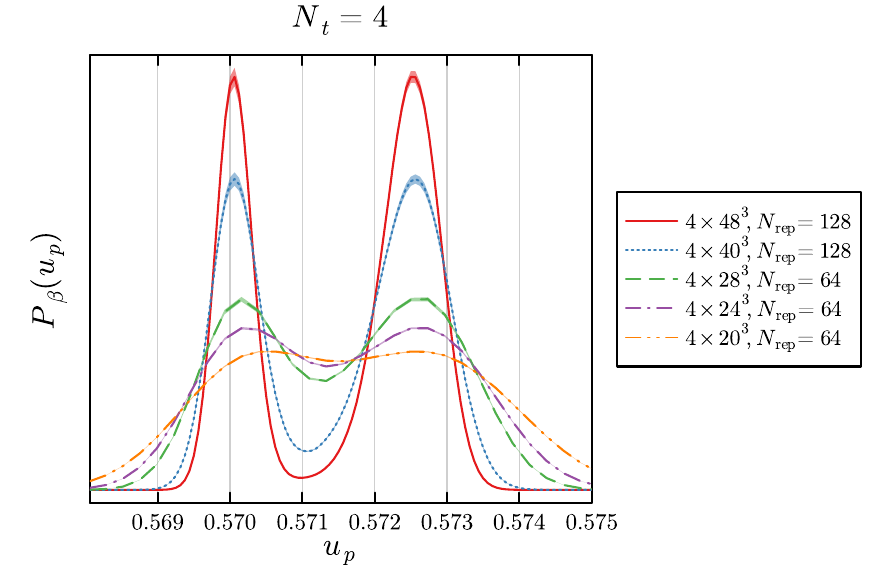}
    \includegraphics[scale=0.577]{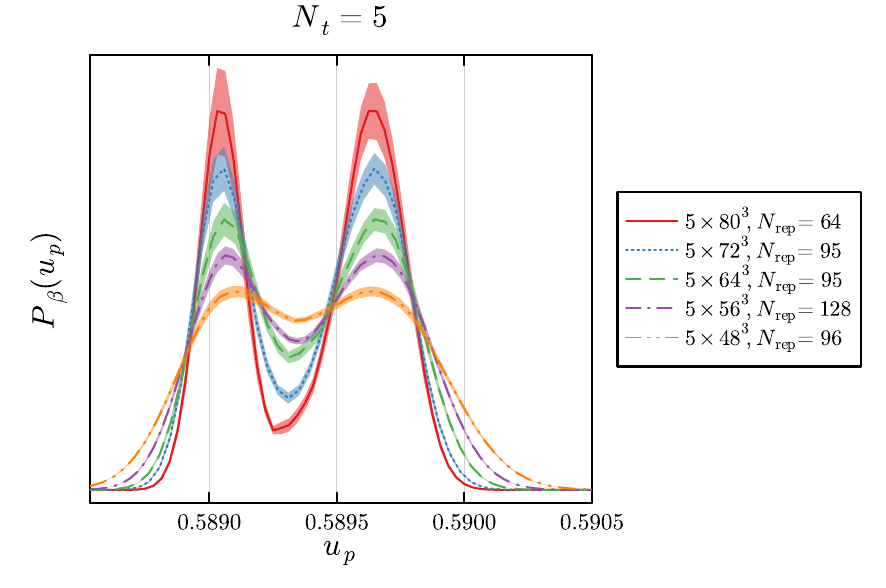}
    \caption{Probability distribution of the plaquette, $P_{\beta}(u_p)$, as defined in Eq.~\eqref{eq:probability_density}, evaluated at the critical coupling, $\beta=\beta_{CV}(P)$, for $N_t=4$ (top panel) and $N_t=5$ (bottom panel), defined by dialing $\beta$  so that the two peaks are of equal height. The probability distribution is normalised so that $\int {\rm d} u_p P_\beta(u_p)=1$.}
    \label{fig:probability_density}
\end{figure}

\begin{figure}[t]
    \centering
    \includegraphics[scale=0.577]{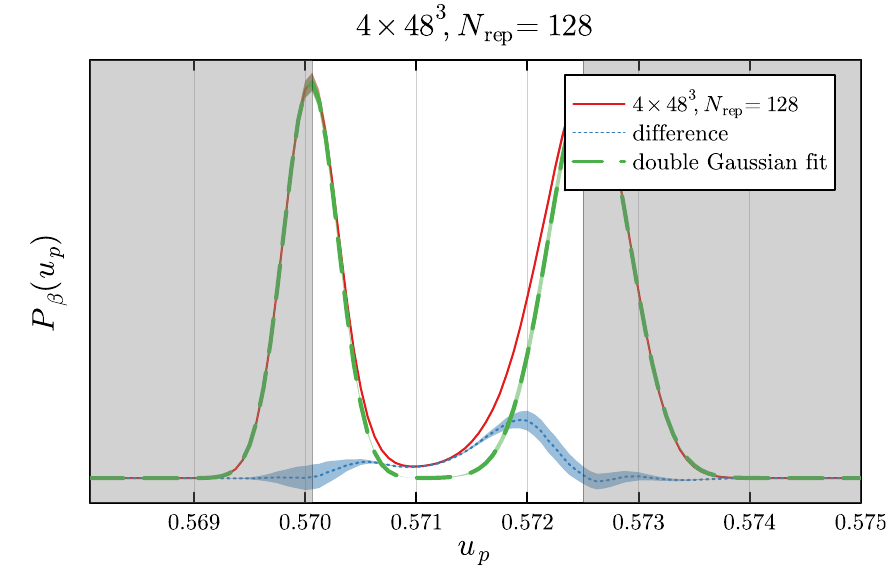}
    \includegraphics[scale=0.577]{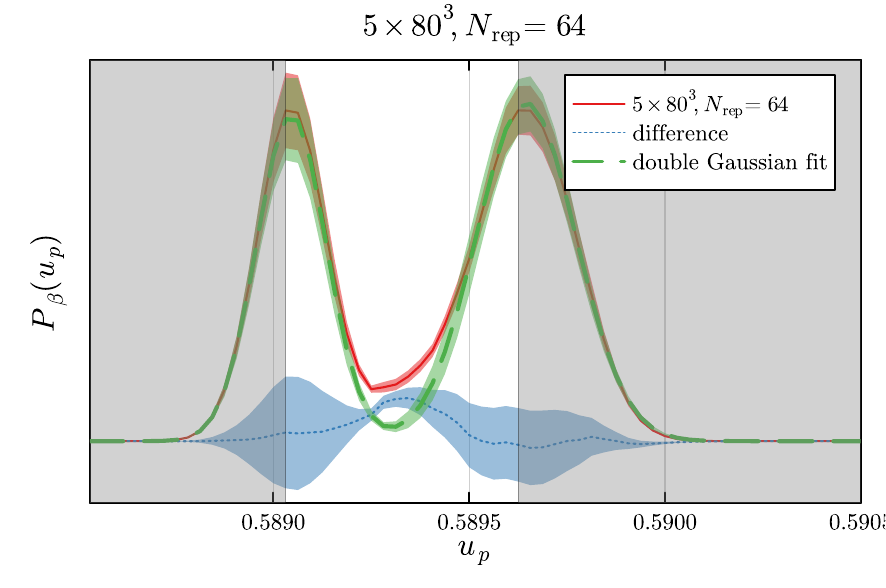}
    \caption{Fit of the plaquette distribution, $P_{\beta}(u_p)$ (blue), to a bimodal Gaussian distribution (green), for the critical value of the coupling, $\beta=\beta_{CV}(P)$, and for the largest available volumes with $N_t=4$ (top panel)  and $N_t=5$ (bottom). The gray shaded region highlights the region of the data used in the fitting procedure. We observe good agreement with the expected Gaussian behaviour for large and small values of the average plaquette, while also detecting clear evidence of the effect of interfaces in the region between the peaks, as evidenced by the non-vanishing difference between measurements and fit (orange).}
    \label{fig:plaquette_dist_fits}
\end{figure}

In Fig.~\ref{fig:interval_comparison}, we show a representative example of the comparison between the values of $a^{(n)}$ obtained with three different choices of the size of the energy intervals, $\Delta_E/2$, holding fixed the lattice parameters $N_t=5$ and $N_s=56$. We find good agreement between results derived with the  interval sizes studied here. We take this as indication that the interval sizes used in generating the measurements presented in this paper are sufficiently small that systematic effects connected to this choice are negligible. 

A common feature of Figs.~\ref{fig:an_volumes} and~\ref{fig:interval_comparison} is that $a^{(n)}$ is not invertible as a function of $u_p$, yet it is unique. To verify that this is a dynamical feature, and not the product of the algorithm we use, we monitored the evolution of the NR and RM steps, and found that,  for this system, independently of the starting point of the algorithm, all solutions converge to a unique value of $a^{(n)}$, for each choice of energy interval. This can be contrasted with some of the findings in the literature on holographic studies of phase transitions (see, for example Fig.~6  in Ref.~\cite{Bea:2021zol}) which expose the possibility that, in some regions of parameter space, multiple values of the temperature (related to $a^{(n)}$) correspond to the same energy. If that were the case in realistic theories, under special conditions the out-of-equilibrium dynamics in proximity of the transition, and the process of bubble nucleation,  might appear quite different from the standard paradigm, with potentially interesting implications for the generation of gravitational  waves. We uncovered no evidence of such phenomena in our current measurements.

\begin{figure}[t]
    \centering
    \includegraphics[scale=0.577]{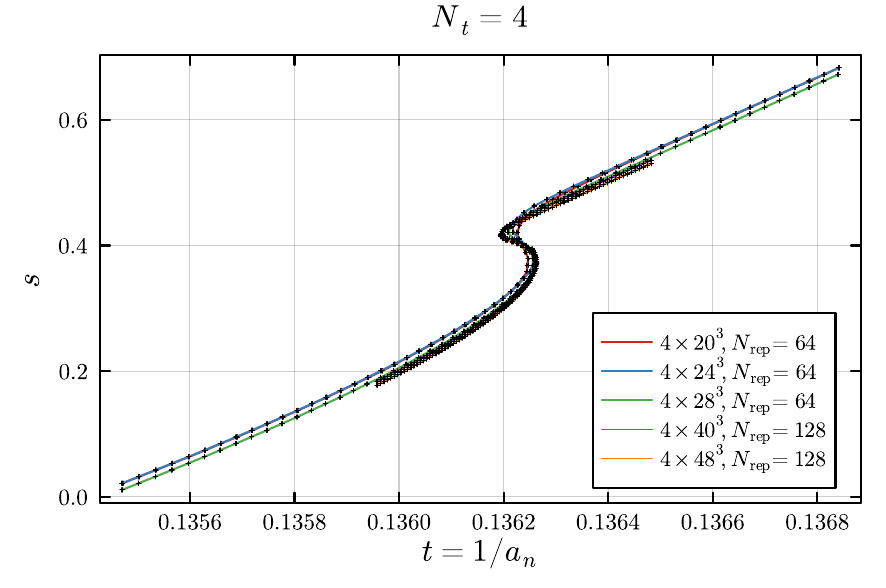}
    \includegraphics[scale=0.577]{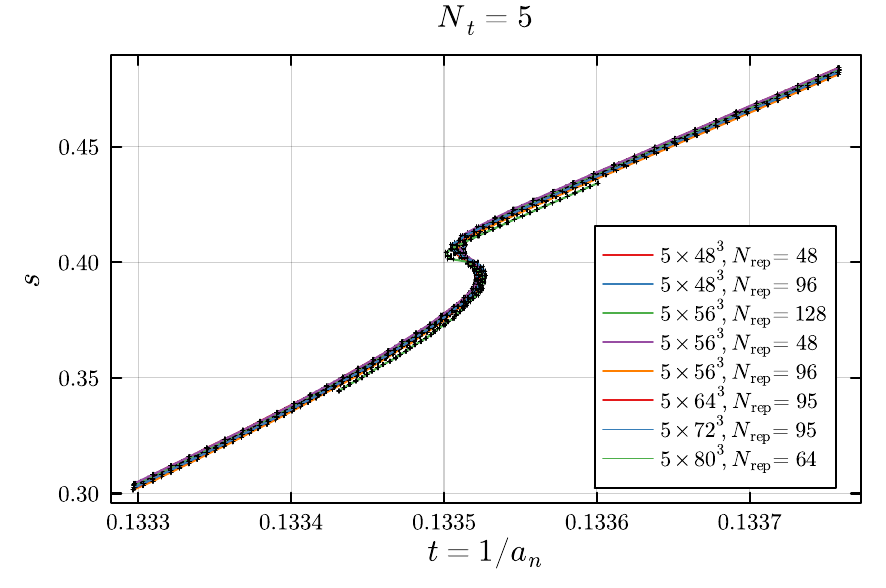}
    \caption{The entropy, $s = \log(\rho)$, measured for $N_t=4$ (top panel) and $N_t=5$ (bottom), for all available calculations. We fix the unknown constant, $c^{(0)}$, in Eq.~\eqref{eq:llr_approx}, and hence an additive constant in $s$,  by requiring that the entropy is positive for all temperatures considered, and furthermore that the value of the entropy evaluated at criticality, on the unstable branch, be the same for all calculations.}
    \label{fig:entropy}
\end{figure}

In Fig.~\ref{fig:probability_density}, we show the ($\beta$-dependent) probability distribution of the average plaquette, $P_{\beta}(u_p)$, tuned to the critical coupling, $\beta=\beta_{CV}(P)$, obtained by requiring that the two peaks of the probability distribution are of equal height. 
The expected double-peaked structure is clearly visible, for both $N_t=4$ and $N_t=5$,  for all available lattice volumes. In the $N_t=5$ case, we were compelled  to use substantially larger lattices and aspect ratios than in the case of $N_t=4$, in order to resolve this feature, and yet the separation of the peaks is less pronounced at the spatial volumes available.

\begin{figure}[t]
    \centering
    \includegraphics[scale=0.577]{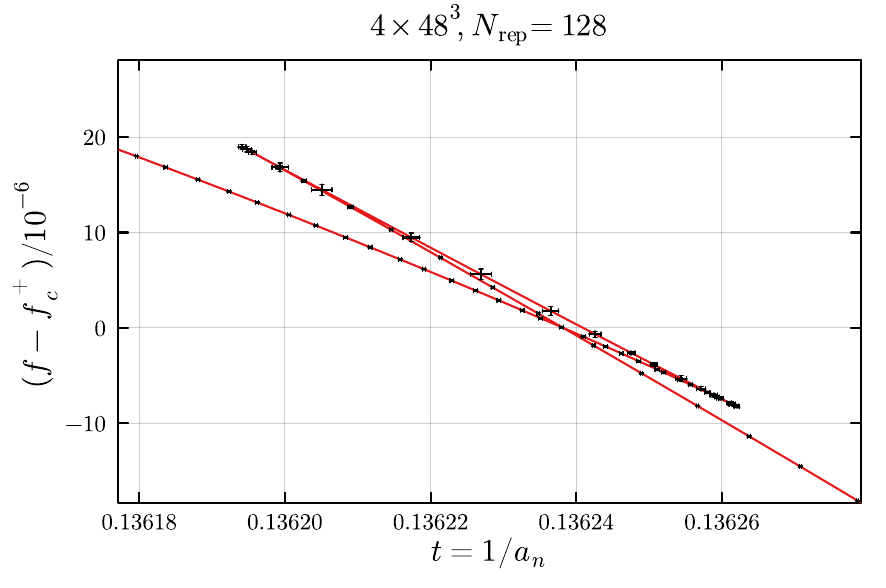}
    \includegraphics[scale=0.577]{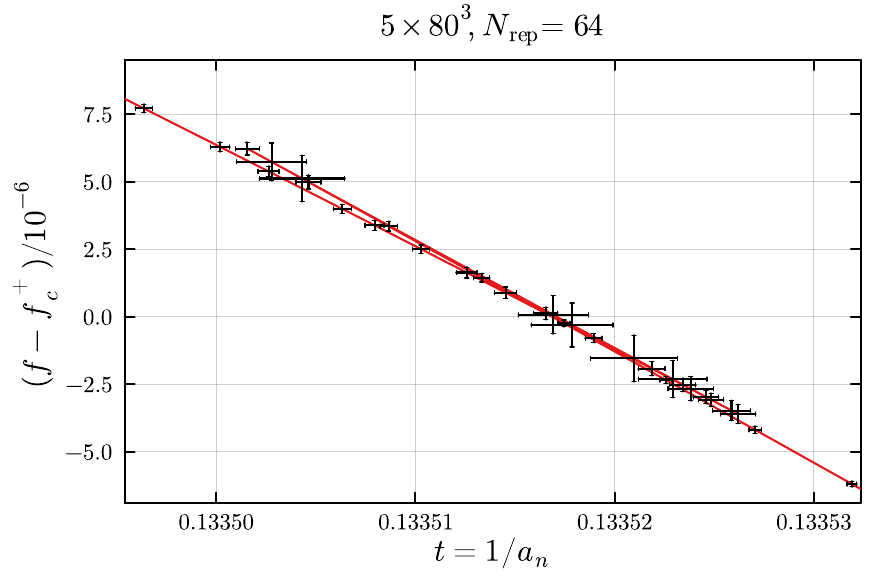}
    \caption{Subtracted free energy density, $f-f_c^+$, determined as the Legendre transform of the internal energy, $E$. We fix the unknown constant in the entropy, $s$, by requiring that the entropy remains positive for all temperatures, and agrees across all different volumes when computed at criticality, on the unstable branch. We find the expected swallow-tail behaviour.}
    \label{fig:swallow_tail}
\end{figure}

In the two panels of Fig.~\ref{fig:plaquette_dist_fits}, we show our measurements of $P_{\beta}(u_p)$ in the largest available spatial volumes, for $N_t=4$ and $N_t=5$, respectively. We fit the probability, $P_{\beta}(u_p)$, computed with $\beta=\beta_{CV}(P)$, to a bimodal Gaussian distribution. If there were no interfaces between the two phases, the plaquette distribution would be given by two Gaussians of equal height, and hence be well reproduced by the bimodal fit. This behavior assumes that there are two independent subsystems corresponding to the two phases---see the discussion in Refs.~\cite{Lucini:2005vg} and \cite{Bennett:2024bhy}. We exclude the energy range between the two peaks from the fit and find that our data fit nicely a Gaussian behaviour, as expected in the absence of interfaces. In the central region, we clearly see the effects of an emerging interface. To help  visualize this qualitative effect, the figure shows also the difference between our data and the fit.
This figure provides clear evidence of a non-vanishing deviation from the Gaussian behaviour in the region between the two peaks, which  encodes the physics of the mixed phases. In particular, for suitably large volumes this study would allow us to measure the surface tension, as anticipated. 

\begin{figure}[t]
    \centering
    \includegraphics[scale=0.577]{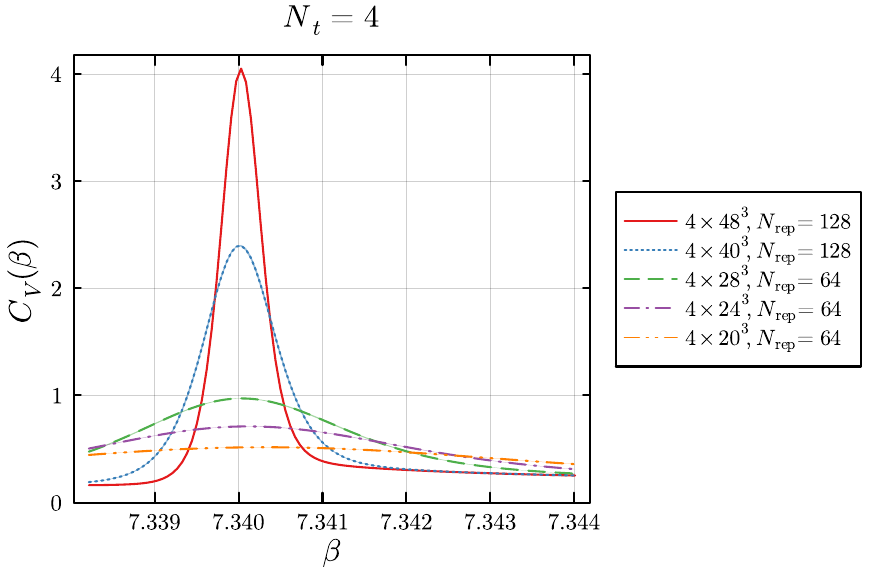}
    \includegraphics[scale=0.577]{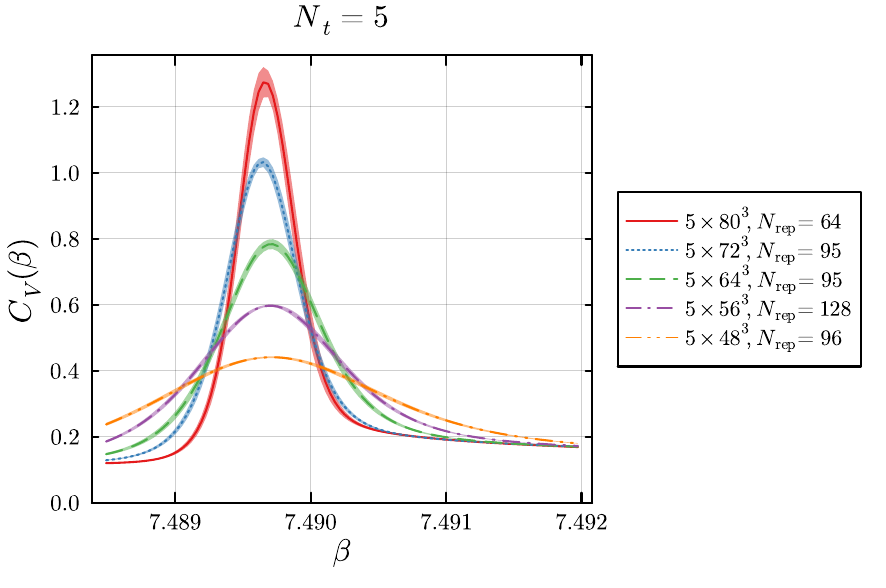}
    \caption{Specific heat, $C_V(\beta)$, for $N_t=4$ (top panel) and $N_t=5$ (bottom). The peaks at $\beta=\beta_{CV}(C_V)$ scale as we approach the thermodynamic limit. We observe that the maximum at $N_t=5$ is lower for the same aspect ratio $N_s/N_t$.}
    \label{fig:specific_heat}
\end{figure}

\begin{figure}[t]
    \centering
    \includegraphics[scale=0.577]{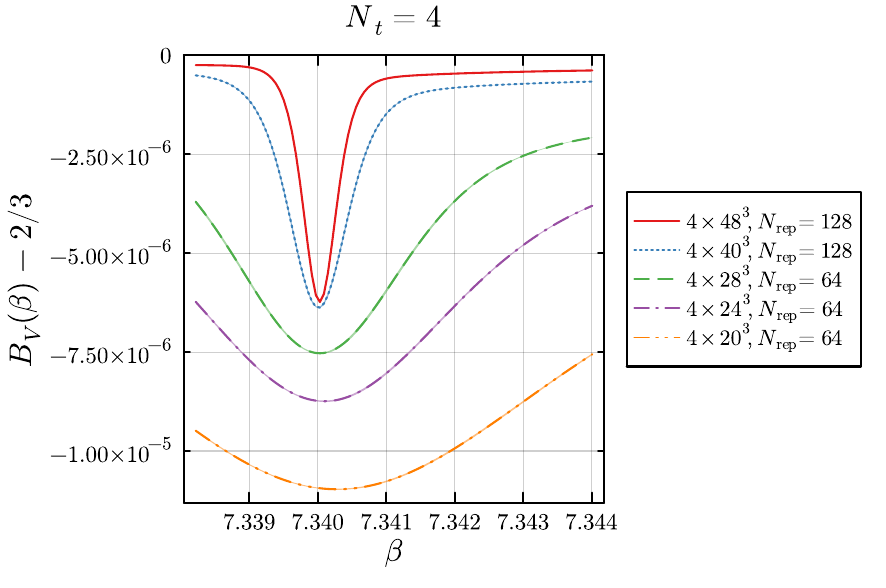}
    \includegraphics[scale=0.577]{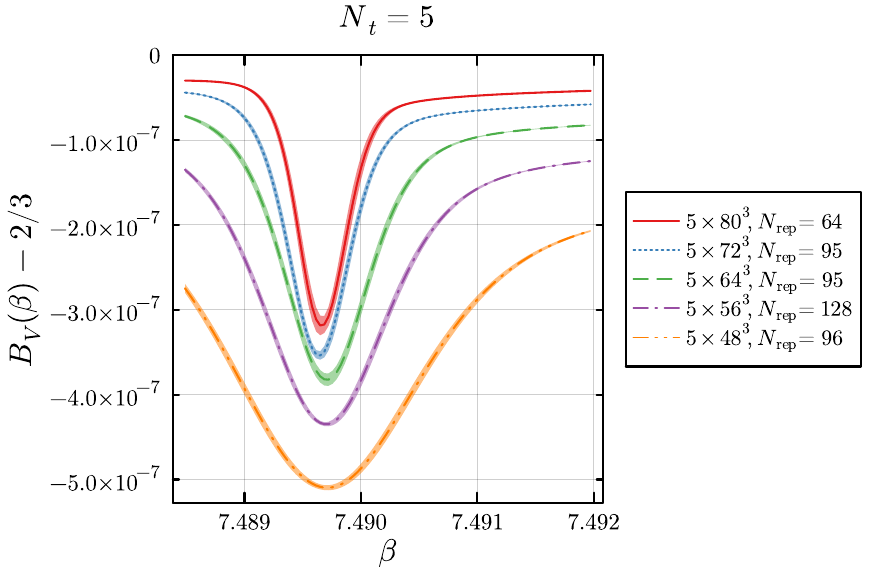}
    \caption{Deviation of the Binder cumulant from two thirds, $B_V(\beta)-2/3$, for $N_t=4$ (top panel) and $N_t=5$ (bottom). We observe a minimum at the critical coupling $\beta_{CV}(B_V)$. }    \label{fig:binder_cumulant}
\end{figure}

\subsection{Thermodynamic properties}

Within the LLR approach, we can determine the entropy only up to an additive constant, given by $c^{(0)}$ in Eq.~\eqref{eq:llr_approx}.
In principle, this constant should be fixed by imposing the third law of thermodynamics, but as anticipated we can only consider temperatures around the phase transition, for this work, and not the low temperatures entering the third law. 
We fix the unknown constant in the entropy by requiring that $s$ be positive for all temperatures considered. Furthermore,
we require that when evaluating the entropy at criticality, on the unstable branch in the regime of phase co-existence,
we obtain the same result for all volumes studied here. 

We show the entropy used in the determination of the free energy in Fig.~\ref{fig:entropy}.
We determine the free energy as the Legendre transform of the internal energy, $E$, and subtract its value at the point where the two metastable branches cross. 
In Fig.~\ref{fig:swallow_tail}, we depict the free energy density, $f=\frac{a^4}{\tilde{V}}F$,  as a function of the micro-canonical temperature, $t=1/a_n$. We observe the appearance of the characteristic swallow-tail shape that accompanies the phenomena of phase coexistence and metastability.
If we were able to probe the system at lower temperatures and fix the unknown constant, in Eq.~\eqref{eq:free_energy}, through the third law of thermodynamics, $\lim_{t \to 0}s=0$, the slope of the free energy would be expected to be steeper.

As a further consistency check, and to assess the size of methodological systematics, we also determine the critical coupling in two more conventional ways, by examining the specific heat (which has a maximum at $\beta = \beta_{CV}(C_V)$), and the Binder cumulant (which has a minimum at $\beta = \beta_{CV}(B_V)$), using Eqs.~\eqref{eq:specific_heat} and~\eqref{eq:binder_cumulant}. We show the specific heat around the critical coupling in Fig.~\ref{fig:specific_heat}, and the Binder cumulant in Fig.~\ref{fig:binder_cumulant}.

\begin{table}
    \centering
    \caption{Three complementary determinations of the critical coupling of the $Sp(4)$ Yang-Mills theory, obtained from different observables: the plaquette distribution, $P_\beta(u_p)$, the specific heat, $C_V(\beta)$, and the Binder cumulant, $B_V(\beta)$. We report our best measurements for  all available lattices.
    For comparison, we report both results for $N_t=4$, which were first presented in Ref.~\cite{Bennett:2024bhy}, and $N_t=5$, original to this work.}
\begin{tabular}{|c|c|c|c|c|c|} \hline
$N_t$ & $N_s$ & $N_{\rm rep}$ & $\beta_{CV}(P)$ & $\beta_{CV}(C_V)$ & $\beta_{CV}(B_V)$ \\ \hline \hline
5 & 48 & 48 & 7.489685(36) & 7.489735(34) & 7.489733(34) \\
5 & 48 & 96 & 7.489663(33) & 7.489707(29) & 7.489705(28) \\
5 & 56 & 128 & 7.489696(22) & 7.489695(20) & 7.489694(20) \\
5 & 56 & 48 & 7.489678(32) & 7.489679(29) & 7.489680(30) \\
5 & 56 & 96 & 7.489678(25) & 7.489680(24) & 7.489681(24) \\
5 & 64 & 95 & 7.489710(24) & 7.489701(21) & 7.489700(21) \\
5 & 72 & 95 & 7.489656(18) & 7.489637(18) & 7.489637(18) \\
5 & 80 & 64 & 7.489690(23) & 7.489666(21) & 7.489666(21) \\
\hline \hline 
4 & 20 & 64 & 7.340131(20) & 7.340317(18) & 7.340282(18) \\
4 & 24 & 64 & 7.340047(15) & 7.340110(14) & 7.340096(15) \\
4 & 28 & 64 & 7.340056(17) & 7.340035(17) & 7.340024(17) \\
4 & 40 & 128 & 7.3400783(97) & 7.3400115(97) & 7.340009(10) \\
4 & 48 & 128 & 7.3400844(57) & 7.3400287(54) & 7.3400275(54) \\
\hline \hline
\end{tabular}

    \label{tab:beta_critical}
\end{table}

\begin{figure}[t]
    \centering
    \includegraphics[scale=0.577]{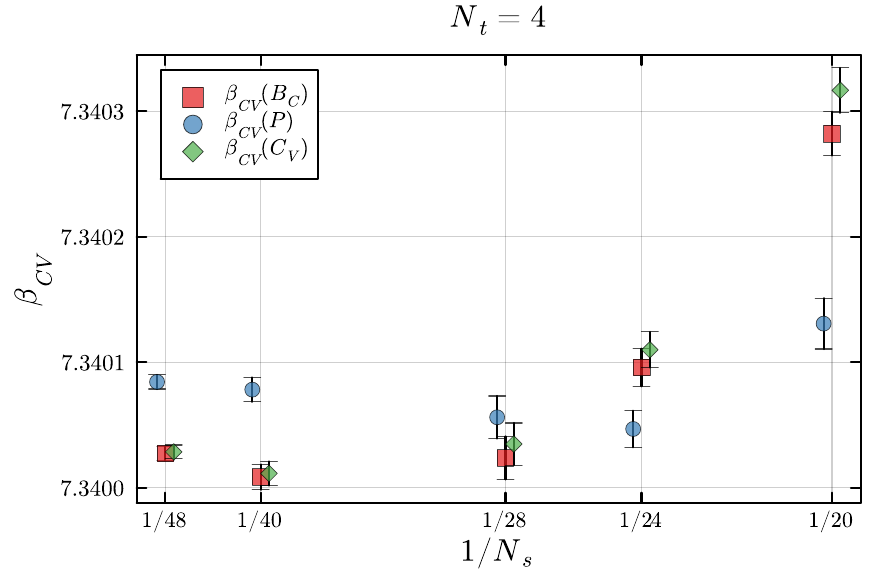}
    \includegraphics[scale=0.577]{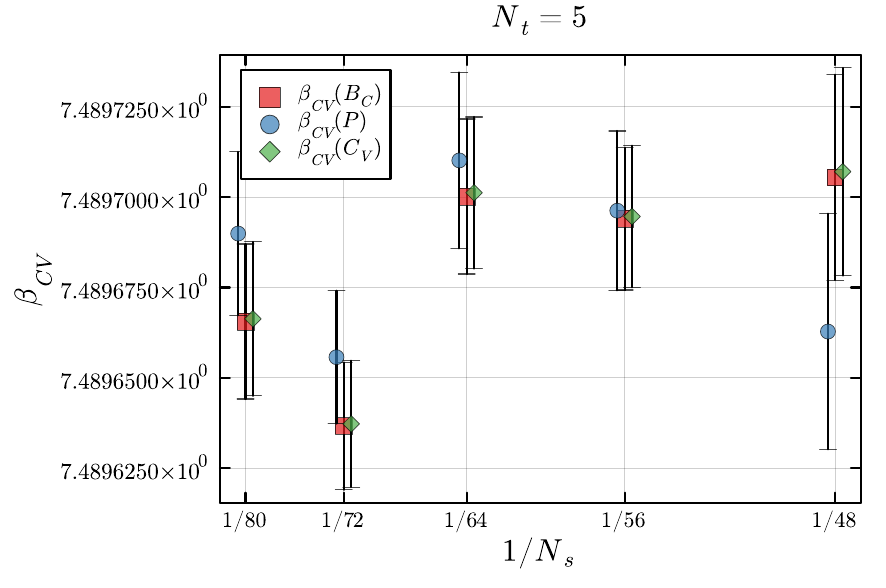}
    \caption{Plot of the data presented in Tab.~\ref{tab:beta_critical}.
    For comparison, we show the critical couplings for both $N_t=4$, which were first presented in Ref.~\cite{Bennett:2024bhy}, and $N_t=5$, original to this work.}
    \label{fig:beta_critical}
\end{figure}

We determine the critical coupling, $\beta_{CV}(C_V)$ ($\beta_{CV}(B_V)$), by identifying the maximum (minimum) of the respective cumulant. 
We show the numerical values of $\beta_{CV}$ obtained at different lattice volumes and using different observables in Tab.~\ref{tab:beta_critical} and Fig.~\ref{fig:beta_critical}. For the measurements with $N_t=4$, these results have already been published in Ref.~\cite{Bennett:2024bhy}, and we observe the presence of statistically significant deviations between the determinations arising from the probability distribution and the specific heat as well as the Binder cumulant. We do not observe such an effect for our new measurements at $N_t=5$ within current uncertainties. 

\begin{figure}[t]
    \centering
    \includegraphics[scale=0.577]{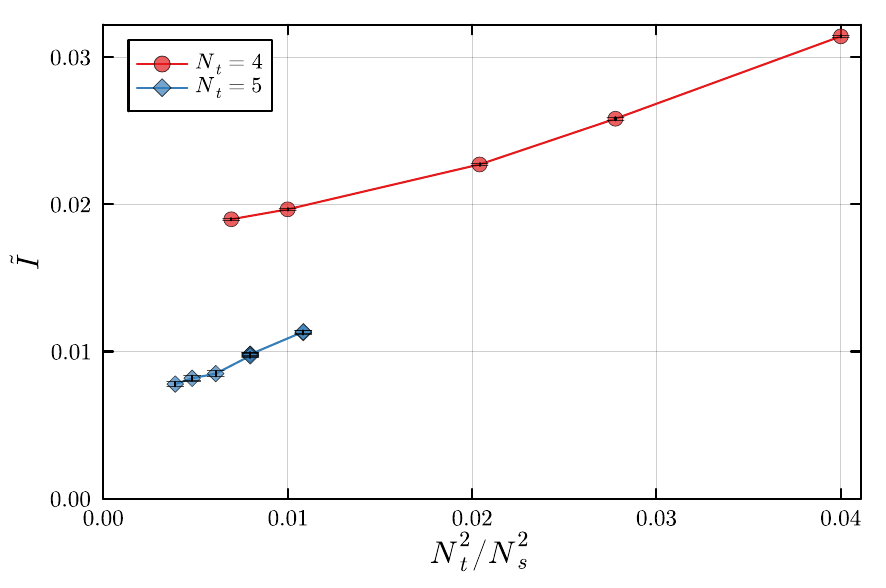}
    \caption{Our numerical results for the term, $\tilde{I}$, giving rise to the dimensionless surface tension, $\sigma_{cd}/T_c^3$, in the limit of vanishing inverse aspect ratio, $N_t/N_s \to 0$, for all available lattices.}
    \label{fig:surface_tension}
\end{figure}

In Fig.~\ref{fig:surface_tension}, we plot $\tilde{I}$, computed according to Eq.~\eqref{eq:surface_tension},
for all available ensembles, including both the cases of
$N_t=4$ and $N_t=5$. We find that the term $\tilde{I}$ is strongly suppressed for $N_t=5$, in comparison to $N_t=4$, for all available 
choices of aspect ratio, $N_t/N_s$, and hence that this quantity is, with current lattices, affected by large lattice artifacts.
It is likely that finer lattices at $N_t \geq 6$ are required to perform the full continuum limit. Nevertheless, the very fact that 
we could extract this quantity brings our current understanding of the surface tensions in the $Sp(4)$ Yang-Mills theory close to the state of the art for other theories, in particular those with $SU(N_c)$ gauge group---see, e.g., Refs.~\cite{deForcrand:2004jt,Lucini:2005vg, Rindlisbacher:2025dqw} and references therein, in particular Refs.~\cite{Iwasaki:1993qu,Grossmann:1992dy,Beinlich:1996xg}. Our results for $\hat I$ are roughly of the same size as for $SU(3)$ \cite{Iwasaki:1993qu,Grossmann:1992dy,Beinlich:1996xg,Lucini:2005vg} suggesting that the transition in $Sp(4)$ is a weak first-order transition. 

\section{Conclusion and outlook}
\label{sec:conclusion}

We presented our results for a new set of lattice studies of the finite-temperature, 
confinement-deconfinement phase transition in the $Sp(4)$ Yang-Mills theory.
Using the LLR algorithm to implement the density of states framework, we made the first steps towards 
the continuum limit extrapolation, by providing  an extensive set of results for lattices with 
temporal extent $N_t=5$, that complement earlier measurements for $N_t=4$. Our calculations have been performed 
 for several choices of (large) spatial volumes, so that the thermodynamic (infinite-volume) limit can be approached.
 We have further shown that for our choices of finite discretization of the energy,  systematic effects arising from the
  LLR implementation are negligible. 

We demonstrated  that we can resolve a first-order phase transition effectively with the LLR method on fine lattices. 

We have used significantly larger lattice volumes than has been done in past investigations within  Yang-Mills theories. We found that larger aspect ratios were required to resolve the phase transition.  
We estimated the critical coupling for finite volumes, $\beta_{CV}$, through several independent prescriptions and found that the measurements are robust, being compatible with one another.

In contrast to existing results for $N_t=4$, in the case of $N_t=5$ this analysis suggests that the error budget is dominated by the statistical uncertainties in $a_n$, rather than by methodological systematics. 

We provided a first quantitative assessment of the size of  discretization artifacts present in the determination of the surface tension.  The current measurement of the surface tension 
 sets an upper limit on its true value, that is
important as input into realistic estimates of the GW power spectrum of the continuum theory. This study provides numerical evidence for the need to deploy even larger lattices, in order to approach the continuum limit and provide a high precision measurement of this quantity.

Further progress is achievable in future studies, by considering finer lattices, and hence larger values of $N_t$.
Doing so will  require considering  larger aspect ratios, $N_s/N_t$, which will increase the cost of the calculations, and demand further  algorithm and software development for the 
implementation of the global constraint on the internal energy underpinning the LLR algorithm, in order to optimize  the parallelization of the underlying calculation.

\begin{figure}[t]
    \centering
    \includegraphics[scale=0.577]{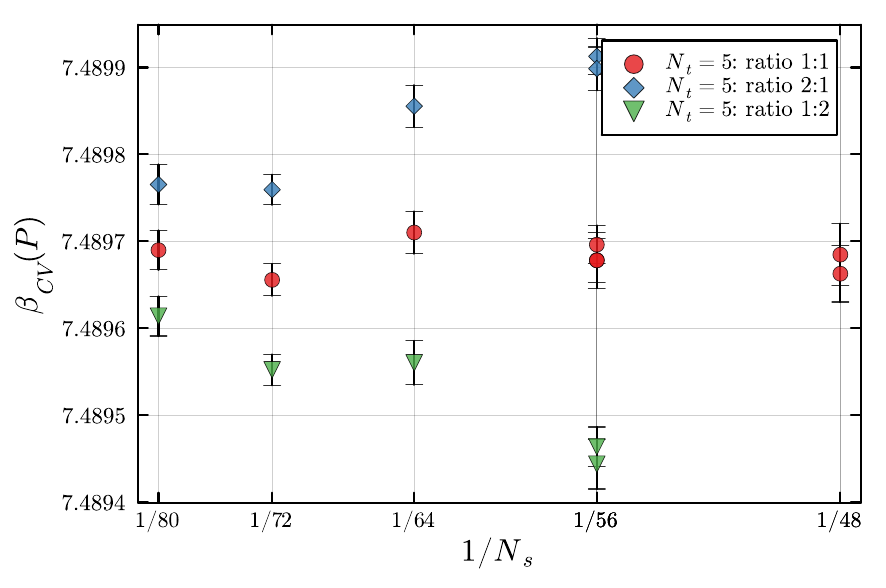}
    \caption{Critical $\beta_{CV}(P)$ as determined (at finite volume) by analysis $P_{\beta}(u_p)$, and dialing $\beta$ to different ratios of peak heights in the plaquette distribution: equal heights (yellow circles) and peak-height-ratio of two to one (blue hexagons) and one to two (green rectangles).}
    \label{fig:beta_c_2:1}
\end{figure}

\begin{acknowledgments}

We would like to thank Stephan J. Huber, David Mateos, and Manuel Reichert for useful discussions, and Frederic D. R. Bonnet for help in benchmarking our code.

The work of E.B. and B.L. is supported in part by the EPSRC ExCALIBUR programme ExaTEPP (project EP/X017168/1). The work of E.B. has also been supported by the UKRI Science and Technology Facilities Council (STFC) Research Software Engineering Fellowship EP/V052489/1. The work of E.B., B.L., and M.P. has been supported in part by the STFC Consolidated Grant No. ST/X000648/1. The work of B.L. is supported in part by the STFC Consolidator Grant No. ST/X00063X/1. The work of B.L. and M.P. has been supported in part by the STFC  Consolidated Grant No. ST/T000813/1.
B.L. and M.P. received funding from the European Research Council (ERC) under the European Union’s Horizon 2020 research and innovation program under Grant Agreement No.~813942. 
D.M. has been supported in part by a studentship awarded by the Data Intensive Centre for Doctoral Training,  funded by the STFC grant ST/P006779/1.
D.V. is supported in part by STFC under Consolidated Grant No.~ST/X000680/1.
F.Z. is supported by the STFC Consolidated Grant No.~ST/X000648/1.

{\bf High performance computing}---This work used the DiRAC Data Intensive service (CSD3) at the University of Cambridge,  the DiRAC Data Intensive service (DIaL3) at the University of Leicester and the DiRAC Extreme Scaling service (Tursa) at the University of Edinburgh, managed respectively by the University of Cambridge University Information Services, the University of Leicester Research Computing Service and by EPCC on behalf of the STFC DiRAC HPC Facility (www.dirac.ac.uk). The DiRAC service at Cambridge, Leicester, and Edinburgh are funded by BEIS, UKRI and STFC capital funding and STFC operations grants. DiRAC is part of the UKRI Digital Research Infrastructure.

This work was supported by the Supercomputer Fugaku Start-up Utilization Program of RIKEN. This work used computational resources of the supercomputer Fugaku provided by RIKEN through the HPCI System Research Project (Project ID: hp230397).

Numerical simulations have been performed on the Swansea SUNBIRD cluster (part of the Supercomputing Wales project). The Swansea SUNBIRD system is part funded by the European Regional Development Fund (ERDF) via Welsh Government.

{\bf Research Software Availability statement}---The workflow used to analyse these data is available at Ref.~\cite{analysis_release}. The modified HiRep code with support for the LLR is available at \cite{mason_HiRep_LLR_v1.1.0}. It is based upon \cite{HiRepSUN,HiRepSpN}.

{\bf Research Data Availability Statement}---The raw data generated in support of this work, and processed data derived from it, are available in machine-readable format at Ref.~\cite{data_release}. See also Ref.~\cite{Bennett:2025neg}, for a description of our approach to reproducibility and open science.

{\bf Open Access Statement}---For the purpose of open access, the authors have applied a Creative Commons  Attribution (CC BY) license to any Author Accepted Manuscript version arising.

\end{acknowledgments}

\appendix

\section{More on determining \texorpdfstring{$\beta_{CV}(P)$}{beta\_c}}\label{sec:peak_ratios}

In Sec.~\ref{sec:results}, we provided a determination of the critical coupling, $\beta_{CV}(P)$, obtained by requiring that the probability distribution of the  plaquette, $P_{\beta}(u_p)$, displays  two peaks of  equal heights.
In this Appendix we consider a modified prescription, in which we require that one of the peaks is twice the height of the other.
In part, this is motivated by the suggestion that, in the presence of multiple choices of vacuum in the one of the phases, 
one might require that such mutiplicity be taken into account~\cite{Janke2003}. 
We hence examine how the resulting measurements  of $\beta_{CV}(P)$, emerging from these three alternative prescriptions, approach  the infinite volume limit. 

We show the results of this analysis in Fig.~\ref{fig:beta_c_2:1}. Finite volume corrections are much more pronounced in the case of prescriptions based on tuning $P_{\beta}(u_p)$ to display peaks with unequal heights. Thus, in the main body of the paper, we restrict ourselves to the definition of $\beta_{CV}(P)$ adopted in from Sec.~\ref{sec:results}. 

\bibliographystyle{JHEP}
\bibliography{ref}

@article{Lucini:2023irm,
    author = "Lucini, Biagio and Mason, David and Piai, Maurizio and Rinaldi, Enrico and Vadacchino, Davide",
    title = "{First-order phase transitions in Yang-Mills theories and the density of state method}",
    eprint = "2305.07463",
    archivePrefix = "arXiv",
    primaryClass = "hep-lat",
    reportNumber = "RIKEN-iTHEMS-Report-23 ET-0164A-23",
    doi = "10.1103/PhysRevD.108.074517",
    journal = "Phys. Rev. D",
    volume = "108",
    number = "7",
    pages = "074517",
    year = "2023"
}

@article{Bea:2024bls,
    author = "Bea, Yago and Giliberti, Mauro and Mateos, David and Sanchez-Garitaonandia, Mikel and Serantes, Alexandre and Zilh{\~a}o, Miguel",
    title = "{Bubble dynamics in a QCD-like phase diagram}",
    eprint = "2412.09588",
    archivePrefix = "arXiv",
    primaryClass = "hep-th",
    month = "12",
    year = "2024"
}

@article{Bea:2024bxu,
    author = "Bea, Yago and Casalderrey-Solana, Jorge and Mateos, David and Sanchez-Garitaonandia, Mikel",
    title = "{Hydrodynamics of Relativistic Superheated Bubbles}",
    eprint = "2406.14450",
    archivePrefix = "arXiv",
    primaryClass = "hep-th",
    reportNumber = "CPHT-RR050.062024",
    month = "6",
    year = "2024"
}

@article{Bea:2024xgv,
    author = "Bea, Yago and Jimenez, Raul and Mateos, David and Liu, Shuheng and Protopapas, Pavlos and Taranc{\'o}n-{\'A}lvarez, Pedro and Tejerina-P{\'e}rez, Pablo",
    title = "{Gravitational duals from equations of state}",
    eprint = "2403.14763",
    archivePrefix = "arXiv",
    primaryClass = "hep-th",
    doi = "10.1007/JHEP07(2024)087",
    journal = "JHEP",
    volume = "07",
    pages = "087",
    year = "2024"
}

@article{Bea:2022mfb,
    author = "Bea, Yago and Casalderrey-Solana, Jorge and Giannakopoulos, Thanasis and Jansen, Aron and Mateos, David and Sanchez-Garitaonandia, Mikel and Zilh{\~a}o, Miguel",
    title = "{Holographic bubbles with Jecco: expanding, collapsing and critical}",
    eprint = "2202.10503",
    archivePrefix = "arXiv",
    primaryClass = "hep-th",
    doi = "10.1007/JHEP09(2022)008",
    journal = "JHEP",
    volume = "09",
    pages = "008",
    year = "2022",
    note = "[Erratum: JHEP 03, 225 (2023)]"
}

@article{Bea:2021zol,
    author = "Bea, Yago and Casalderrey-Solana, Jorge and Giannakopoulos, Thanasis and Jansen, Aron and Krippendorf, Sven and Mateos, David and Sanchez-Garitaonandia, Mikel and Zilh{\~a}o, Miguel",
    title = "{Spinodal Gravitational Waves}",
    eprint = "2112.15478",
    archivePrefix = "arXiv",
    primaryClass = "hep-th",
    month = "12",
    year = "2021"
}

@book{Kalikmanov2013,
    author = "Kalikmanov, V. I.",
    title = "{Nucleation Theory, Lecture Notes in Physics, Volume 860}",
    doi = " 10.1007/978-90-481-3643-8",
    isbn = "978-90-481-3642-1",
    publisher = "Springer Science+Business Media",
    address = "Dordrecht, NL",
    year = "2013"
}

@article{afzal2023nanograv,
  title={The nanograv 15 yr data set: Search for signals from new physics},
  author={Afzal, Adeela and Agazie, Gabriella and Anumarlapudi, Akash and Archibald, Anne M and Arzoumanian, Zaven and Baker, Paul T and B{\'e}csy, Bence and Blanco-Pillado, Jose Juan and Blecha, Laura and Boddy, Kimberly K and others},
  journal={The Astrophysical Journal Letters},
  volume={951},
  number={1},
  pages={L11},
  year={2023},
  publisher={IOP Publishing}
}

@article{Pomper:2024otb,
    author = "Pomper, Joachim and Kulkarni, Suchita",
    title = "{Low energy effective theories of composite dark matter with real representations}",
    eprint = "2402.04176",
    archivePrefix = "arXiv",
    primaryClass = "hep-ph",
    month = "2",
    year = "2024"
}

@article{Cacciapaglia:2023kat,
    author = "Cacciapaglia, Giacomo and Cheong, Dhong Yeon and Deandrea, Aldo and Isnard, Wanda and Park, Seong Chan",
    title = "{Composite hybrid inflation: dilaton and waterfall pions}",
    eprint = "2307.01852",
    archivePrefix = "arXiv",
    primaryClass = "hep-ph",
    doi = "10.1088/1475-7516/2023/10/063",
    journal = "JCAP",
    volume = "10",
    pages = "063",
    year = "2023"
}

@article{Ferrante:2023bcz,
    author = "Ferrante, Steven and Ismail, Ameen and Lee, Seung J. and Lee, Yunha",
    title = "{Forbidden conformal dark matter at a GeV}",
    eprint = "2308.16219",
    archivePrefix = "arXiv",
    primaryClass = "hep-ph",
    doi = "10.1007/JHEP11(2023)186",
    journal = "JHEP",
    volume = "11",
    pages = "186",
    year = "2023"
}

@article{Appelquist:2024koa,
    author = "Appelquist, Thomas and Ingoldby, James and Piai, Maurizio",
    title = "{Dilaton forbidden dark matter}",
    eprint = "2404.07601",
    archivePrefix = "arXiv",
    primaryClass = "hep-ph",
    reportNumber = "IPPP/24/17",
    doi = "10.1103/PhysRevD.110.035013",
    journal = "Phys. Rev. D",
    volume = "110",
    number = "3",
    pages = "035013",
    year = "2024"
}

@article{Bruno:2024dha,
    author = "Bruno, Mattia and Forzano, Niccol{\`o} and Panero, Marco and Smecca, Antonio",
    title = "{Thermal evolution of dark matter in the early universe from a symplectic glueball model}",
    eprint = "2410.17122",
    archivePrefix = "arXiv",
    primaryClass = "hep-ph",
    month = "10",
    year = "2024"
}

@article{Sakharov:1967dj,
    author = "Sakharov, A. D.",
    title = "{Violation of CP Invariance, C asymmetry, and baryon asymmetry of the universe}",
    doi = "10.1070/PU1991v034n05ABEH002497",
    journal = "Pisma Zh. Eksp. Teor. Fiz.",
    volume = "5",
    pages = "32--35",
    year = "1967"
}

@article{Kajantie:1996mn,
    author = "Kajantie, K. and Laine, M. and Rummukainen, K. and Shaposhnikov, Mikhail E.",
    title = "{Is there a~ hot electroweak phase transition at $m_H \gtrsim m_W$?}",
    eprint = "hep-ph/9605288",
    archivePrefix = "arXiv",
    reportNumber = "CERN-TH-96-126, HD-THEP-96-15, IUHET-333",
    doi = "10.1103/PhysRevLett.77.2887",
    journal = "Phys. Rev. Lett.",
    volume = "77",
    pages = "2887--2890",
    year = "1996"
}

@article{Karsch:1996yh,
    author = "Karsch, F. and Neuhaus, T. and Patkos, A. and Rank, J.",
    editor = "Bernard, C. and Golterman, M. and Ogilvie, M. and Potvin, J.",
    title = "{Critical Higgs mass and temperature dependence of gauge boson masses in the SU(2) gauge Higgs model}",
    eprint = "hep-lat/9608087",
    archivePrefix = "arXiv",
    reportNumber = "FSU-SCRI-96C-79",
    doi = "10.1016/S0920-5632(96)00736-0",
    journal = "Nucl. Phys. B Proc. Suppl.",
    volume = "53",
    pages = "623--625",
    year = "1997"
}

@article{Gurtler:1997hr,
    author = "Gurtler, M. and Ilgenfritz, Ernst-Michael and Schiller, A.",
    title = "{Where the electroweak phase transition ends}",
    eprint = "hep-lat/9704013",
    archivePrefix = "arXiv",
    reportNumber = "UL-NTZ-10-97, HUB-EP-97-24, DESY-97-086",
    doi = "10.1103/PhysRevD.56.3888",
    journal = "Phys. Rev. D",
    volume = "56",
    pages = "3888--3895",
    year = "1997"
}

@article{Rummukainen:1998as,
    author = "Rummukainen, K. and Tsypin, M. and Kajantie, K. and Laine, M. and Shaposhnikov, Mikhail E.",
    title = "{The Universality class of the electroweak theory}",
    eprint = "hep-lat/9805013",
    archivePrefix = "arXiv",
    reportNumber = "CERN-TH-98-08, NORDITA-98-30-HE",
    doi = "10.1016/S0550-3213(98)00494-5",
    journal = "Nucl. Phys. B",
    volume = "532",
    pages = "283--314",
    year = "1998"
}

@article{Csikor:1998eu,
    author = "Csikor, F. and Fodor, Z. and Heitger, J.",
    title = "{Endpoint of the hot electroweak phase transition}",
    eprint = "hep-ph/9809291",
    archivePrefix = "arXiv",
    reportNumber = "ITP-BUDAPEST-541, KEK-TH-580, KEK-PREPRINT-98-160, MS-TPI-98-16",
    doi = "10.1103/PhysRevLett.82.21",
    journal = "Phys. Rev. Lett.",
    volume = "82",
    pages = "21--24",
    year = "1999"
}

@article{Aoki:1999fi,
    author = "Aoki, Y. and Csikor, F. and Fodor, Z. and Ukawa, A.",
    title = "{The Endpoint of the first order phase transition of the SU(2) gauge Higgs model on a four-dimensional isotropic lattice}",
    eprint = "hep-lat/9901021",
    archivePrefix = "arXiv",
    reportNumber = "ITP-BUDAPEST-547, UTCCP-P-60, UTHEP-397",
    doi = "10.1103/PhysRevD.60.013001",
    journal = "Phys. Rev. D",
    volume = "60",
    pages = "013001",
    year = "1999"
}

@article{DOnofrio:2015gop,
    author = "D'Onofrio, Michela and Rummukainen, Kari",
    title = "{Standard model cross-over on the lattice}",
    eprint = "1508.07161",
    archivePrefix = "arXiv",
    primaryClass = "hep-ph",
    reportNumber = "HIP-2015-30-TH",
    doi = "10.1103/PhysRevD.93.025003",
    journal = "Phys. Rev. D",
    volume = "93",
    number = "2",
    pages = "025003",
    year = "2016"
}

@article{Laine:1998jb,
    author = "Laine, M. and Rummukainen, K.",
    editor = "DeGrand, Thomas A. and DeTar, Carleton E. and Sugar, R. and Toussaint, D.",
    title = "{What's new with the electroweak phase transition?}",
    eprint = "hep-lat/9809045",
    archivePrefix = "arXiv",
    doi = "10.1016/S0920-5632(99)85017-8",
    journal = "Nucl. Phys. B Proc. Suppl.",
    volume = "73",
    pages = "180--185",
    year = "1999"
}

@article{Morrissey:2012db,
    author = "Morrissey, David E. and Ramsey-Musolf, Michael J.",
    title = "{Electroweak baryogenesis}",
    eprint = "1206.2942",
    archivePrefix = "arXiv",
    primaryClass = "hep-ph",
    reportNumber = "NPAC-12-08",
    doi = "10.1088/1367-2630/14/12/125003",
    journal = "New J. Phys.",
    volume = "14",
    pages = "125003",
    year = "2012"
}

@article{Gould:2022ran,
    author = {Gould, Oliver and G\"uyer, Sinan and Rummukainen, Kari},
    title = "{First-order electroweak phase transitions: A nonperturbative update}",
    eprint = "2205.07238",
    archivePrefix = "arXiv",
    primaryClass = "hep-lat",
    reportNumber = "HIP-2022-10/TH",
    doi = "10.1103/PhysRevD.106.114507",
    journal = "Phys. Rev. D",
    volume = "106",
    number = "11",
    pages = "114507",
    year = "2022"
}

@article{Witten:1984rs,
    author = "Witten, Edward",
    title = "{Cosmic Separation of Phases}",
    reportNumber = "PRINT-84-0400 (IAS,PRINCETON)",
    doi = "10.1103/PhysRevD.30.272",
    journal = "Phys. Rev. D",
    volume = "30",
    pages = "272--285",
    year = "1984"
}

@article{Kamionkowski:1993fg,
    author = "Kamionkowski, Marc and Kosowsky, Arthur and Turner, Michael S.",
    title = "{Gravitational radiation from first order phase transitions}",
    eprint = "astro-ph/9310044",
    archivePrefix = "arXiv",
    reportNumber = "IASSNS-HEP-93-44, FERMILAB-PUB-93-235-A",
    doi = "10.1103/PhysRevD.49.2837",
    journal = "Phys. Rev. D",
    volume = "49",
    pages = "2837--2851",
    year = "1994"
}

@inproceedings{Allen:1996vm,
    author = "Allen, Bruce",
    title = "{The Stochastic gravity wave background: Sources and detection}",
    booktitle = "{Les Houches School of Physics: Astrophysical Sources of Gravitational Radiation}",
    eprint = "gr-qc/9604033",
    archivePrefix = "arXiv",
    reportNumber = "WISC-MILW-96-TH-22",
    pages = "373--417",
    month = "4",
    year = "1996"
}

@article{Schwaller:2015tja,
    author = "Schwaller, Pedro",
    title = "{Gravitational Waves from a Dark Phase Transition}",
    eprint = "1504.07263",
    archivePrefix = "arXiv",
    primaryClass = "hep-ph",
    reportNumber = "CERN-PH-TH-2015-093",
    doi = "10.1103/PhysRevLett.115.181101",
    journal = "Phys. Rev. Lett.",
    volume = "115",
    number = "18",
    pages = "181101",
    year = "2015"
}

@article{Croon:2018erz,
    author = "Croon, Djuna and Sanz, Ver\'onica and White, Graham",
    title = "{Model Discrimination in Gravitational Wave spectra from Dark Phase Transitions}",
    eprint = "1806.02332",
    archivePrefix = "arXiv",
    primaryClass = "hep-ph",
    doi = "10.1007/JHEP08(2018)203",
    journal = "JHEP",
    volume = "08",
    pages = "203",
    year = "2018"
}

@article{Christensen:2018iqi,
    author = "Christensen, Nelson",
    title = "{Stochastic Gravitational Wave Backgrounds}",
    eprint = "1811.08797",
    archivePrefix = "arXiv",
    primaryClass = "gr-qc",
    doi = "10.1088/1361-6633/aae6b5",
    journal = "Rept. Prog. Phys.",
    volume = "82",
    number = "1",
    pages = "016903",
    year = "2019"
}

@article{Seto:2001qf,
    author = "Seto, Naoki and Kawamura, Seiji and Nakamura, Takashi",
    title = "{Possibility of direct measurement of the acceleration of the universe using 0.1-Hz band laser interferometer gravitational wave antenna in space}",
    eprint = "astro-ph/0108011",
    archivePrefix = "arXiv",
    doi = "10.1103/PhysRevLett.87.221103",
    journal = "Phys. Rev. Lett.",
    volume = "87",
    pages = "221103",
    year = "2001"
}

@article{Kawamura:2006up,
    author = "Kawamura, S. and others",
    editor = "Mio, N.",
    title = "{The Japanese space gravitational wave antenna DECIGO}",
    doi = "10.1088/0264-9381/23/8/S17",
    journal = "Class. Quant. Grav.",
    volume = "23",
    pages = "S125--S132",
    year = "2006"
}

@article{Crowder:2005nr,
    author = "Crowder, Jeff and Cornish, Neil J.",
    title = "{Beyond LISA: Exploring future gravitational wave missions}",
    eprint = "gr-qc/0506015",
    archivePrefix = "arXiv",
    doi = "10.1103/PhysRevD.72.083005",
    journal = "Phys. Rev. D",
    volume = "72",
    pages = "083005",
    year = "2005"
}

@article{Corbin:2005ny,
    author = "Corbin, Vincent and Cornish, Neil J.",
    title = "{Detecting the cosmic gravitational wave background with the big bang observer}",
    eprint = "gr-qc/0512039",
    archivePrefix = "arXiv",
    doi = "10.1088/0264-9381/23/7/014",
    journal = "Class. Quant. Grav.",
    volume = "23",
    pages = "2435--2446",
    year = "2006"
}

@article{Harry:2006fi,
    author = "Harry, G. M. and Fritschel, P. and Shaddock, D. A. and Folkner, W. and Phinney, E. S.",
    title = "{Laser interferometry for the big bang observer}",
    doi = "10.1088/0264-9381/23/15/008",
    journal = "Class. Quant. Grav.",
    volume = "23",
    pages = "4887--4894",
    year = "2006",
    note = "[Erratum: Class.Quant.Grav. 23, 7361 (2006)]"
}

@article{Hild:2010id,
    author = "Hild, S. and others",
    title = "{Sensitivity Studies for Third-Generation Gravitational Wave Observatories}",
    eprint = "1012.0908",
    archivePrefix = "arXiv",
    primaryClass = "gr-qc",
    doi = "10.1088/0264-9381/28/9/094013",
    journal = "Class. Quant. Grav.",
    volume = "28",
    pages = "094013",
    year = "2011"
}

@article{Yagi:2011wg,
    author = "Yagi, Kent and Seto, Naoki",
    title = "{Detector configuration of DECIGO/BBO and identification of cosmological neutron-star binaries}",
    eprint = "1101.3940",
    archivePrefix = "arXiv",
    primaryClass = "astro-ph.CO",
    doi = "10.1103/PhysRevD.83.044011",
    journal = "Phys. Rev. D",
    volume = "83",
    pages = "044011",
    year = "2011",
    note = "[Erratum: Phys.Rev.D 95, 109901 (2017)]"
}

@article{Sathyaprakash:2012jk,
    author = "Sathyaprakash, B. and others",
    editor = "Hannam, Mark and Sutton, Patrick and Hild, Stefan and van den Broeck, Chris",
    title = "{Scientific Objectives of Einstein Telescope}",
    eprint = "1206.0331",
    archivePrefix = "arXiv",
    primaryClass = "gr-qc",
    doi = "10.1088/0264-9381/29/12/124013",
    journal = "Class. Quant. Grav.",
    volume = "29",
    pages = "124013",
    year = "2012",
    note = "[Erratum: Class.Quant.Grav. 30, 079501 (2013)]"
}

@article{Thrane:2013oya,
    author = "Thrane, Eric and Romano, Joseph D.",
    title = "{Sensitivity curves for searches for gravitational-wave backgrounds}",
    eprint = "1310.5300",
    archivePrefix = "arXiv",
    primaryClass = "astro-ph.IM",
    doi = "10.1103/PhysRevD.88.124032",
    journal = "Phys. Rev. D",
    volume = "88",
    number = "12",
    pages = "124032",
    year = "2013"
}

@article{Caprini:2015zlo,
    author = "Caprini, Chiara and others",
    title = "{Science with the space-based interferometer eLISA. II: Gravitational waves from cosmological phase transitions}",
    eprint = "1512.06239",
    archivePrefix = "arXiv",
    primaryClass = "astro-ph.CO",
    reportNumber = "DESY-15-246",
    doi = "10.1088/1475-7516/2016/04/001",
    journal = "JCAP",
    volume = "04",
    pages = "001",
    year = "2016"
}

@article{LISA:2017pwj,
    author = "Amaro-Seoane, Pau and others",
    collaboration = "LISA",
    title = "{Laser Interferometer Space Antenna}",
    eprint = "1702.00786",
    archivePrefix = "arXiv",
    primaryClass = "astro-ph.IM",
    month = "2",
    year = "2017"
}

@inbook{Janke2003,
author="Janke, W.",
editor="D{\"u}nweg, Burkhard
and Landau, David P.
and Milchev, Andrey I.",
title="First-Order Phase Transitions",
bookTitle="Computer Simulations of Surfaces and Interfaces",
year="2003",
publisher="Springer Netherlands",
address="Dordrecht",
pages="111--135",
isbn="978-94-010-0173-1",
doi="10.1007/978-94-010-0173-1_6",
url="https://doi.org/10.1007/978-94-010-0173-1_6"
}

@article{LIGOScientific:2016wof,
    author = "Abbott, Benjamin P and others",
    collaboration = "LIGO Scientific",
    title = "{Exploring the Sensitivity of Next Generation Gravitational Wave Detectors}",
    eprint = "1607.08697",
    archivePrefix = "arXiv",
    primaryClass = "astro-ph.IM",
    reportNumber = "LIGO-P1600143",
    doi = "10.1088/1361-6382/aa51f4",
    journal = "Class. Quant. Grav.",
    volume = "34",
    number = "4",
    pages = "044001",
    year = "2017"
}

@article{Isoyama:2018rjb,
    author = "Isoyama, Soichiro and Nakano, Hiroyuki and Nakamura, Takashi",
    title = "{Multiband Gravitational-Wave Astronomy: Observing binary inspirals with a decihertz detector, B-DECIGO}",
    eprint = "1802.06977",
    archivePrefix = "arXiv",
    primaryClass = "gr-qc",
    doi = "10.1093/ptep/pty078",
    journal = "PTEP",
    volume = "2018",
    number = "7",
    pages = "073E01",
    year = "2018"
}

@misc{Baker:2019nia,
    author = "Baker, John and others",
    title = "{The Laser Interferometer Space Antenna: Unveiling the Millihertz Gravitational Wave Sky}",
    eprint = "1907.06482",
    archivePrefix = "arXiv",
    primaryClass = "astro-ph.IM",
    reportNumber = "FERMILAB-PUB-19-436-A",
    month = "7",
    year = "2019"
}

@article{Brdar:2018num,
    author = "Brdar, Vedran and Helmboldt, Alexander J. and Kubo, Jisuke",
    title = "{Gravitational Waves from First-Order Phase Transitions: LIGO as a Window to Unexplored Seesaw Scales}",
    eprint = "1810.12306",
    archivePrefix = "arXiv",
    primaryClass = "hep-ph",
    doi = "10.1088/1475-7516/2019/02/021",
    journal = "JCAP",
    volume = "02",
    pages = "021",
    year = "2019"
}

@article{Reitze:2019iox,
    author = "Reitze, David and others",
    title = "{Cosmic Explorer: The U.S. Contribution to Gravitational-Wave Astronomy beyond LIGO}",
    eprint = "1907.04833",
    archivePrefix = "arXiv",
    primaryClass = "astro-ph.IM",
    reportNumber = "LIGO-P1900316",
    journal = "Bull. Am. Astron. Soc.",
    volume = "51",
    number = "7",
    pages = "035",
    year = "2019"
}

@article{Caprini:2019egz,
    author = "Caprini, Chiara and others",
    title = "{Detecting gravitational waves from cosmological phase transitions with LISA: an update}",
    eprint = "1910.13125",
    archivePrefix = "arXiv",
    primaryClass = "astro-ph.CO",
    reportNumber = "DESY-19-159, IPPP/19/27, HIP-2019-14/TH, MITP/19-066, IFT-UAM/CSIC-19-139",
    doi = "10.1088/1475-7516/2020/03/024",
    journal = "JCAP",
    volume = "03",
    pages = "024",
    year = "2020"
}

@article{Maggiore:2019uih,
    author = "Maggiore, Michele and others",
    title = "{Science Case for the Einstein Telescope}",
    eprint = "1912.02622",
    archivePrefix = "arXiv",
    primaryClass = "astro-ph.CO",
    doi = "10.1088/1475-7516/2020/03/050",
    journal = "JCAP",
    volume = "03",
    pages = "050",
    year = "2020"
}

@article{Huang:2020crf,
    author = "Huang, Wei-Chih and Reichert, Manuel and Sannino, Francesco and Wang, Zhi-Wei",
    title = "{Testing the dark SU(N) Yang-Mills theory confined landscape: From the lattice to gravitational waves}",
    eprint = "2012.11614",
    archivePrefix = "arXiv",
    primaryClass = "hep-ph",
    doi = "10.1103/PhysRevD.104.035005",
    journal = "Phys. Rev. D",
    volume = "104",
    number = "3",
    pages = "035005",
    year = "2021"
}

@article{Halverson:2020xpg,
    author = "Halverson, James and Long, Cody and Maiti, Anindita and Nelson, Brent and Salinas, Gustavo",
    title = "{Gravitational waves from dark Yang-Mills sectors}",
    eprint = "2012.04071",
    archivePrefix = "arXiv",
    primaryClass = "hep-ph",
    doi = "10.1007/JHEP05(2021)154",
    journal = "JHEP",
    volume = "05",
    pages = "154",
    year = "2021"
}

@article{Kang:2021epo,
    author = "Kang, Zhaofeng and Zhu, Jiang and Matsuzaki, Shinya",
    title = "{Dark confinement-deconfinement phase transition: a roadmap from Polyakov loop models to gravitational waves}",
    eprint = "2101.03795",
    archivePrefix = "arXiv",
    primaryClass = "hep-ph",
    doi = "10.1007/JHEP09(2021)060",
    journal = "JHEP",
    volume = "09",
    pages = "060",
    year = "2021"
}

@article{Reichert:2021cvs,
    author = "Reichert, Manuel and Sannino, Francesco and Wang, Zhi-Wei and Zhang, Chen",
    title = "{Dark confinement and chiral phase transitions: gravitational waves vs matter representations}",
    eprint = "2109.11552",
    archivePrefix = "arXiv",
    primaryClass = "hep-ph",
    doi = "10.1007/JHEP01(2022)003",
    journal = "JHEP",
    volume = "01",
    pages = "003",
    year = "2022"
}

@article{Reichert:2022naa,
    author = "Reichert, Manuel and Wang, Zhi-Wei",
    title = "{Gravitational Waves from dark composite dynamics}",
    eprint = "2211.08877",
    archivePrefix = "arXiv",
    primaryClass = "hep-ph",
    doi = "10.1051/epjconf/202227408003",
    journal = "EPJ Web Conf.",
    volume = "274",
    pages = "08003",
    year = "2022"
}

@article{Pasechnik:2023hwv,
    author = "Pasechnik, Roman and Reichert, Manuel and Sannino, Francesco and Wang, Zhi-Wei",
    title = "{Gravitational waves from composite dark sectors}",
    eprint = "2309.16755",
    archivePrefix = "arXiv",
    primaryClass = "hep-ph",
    doi = "10.1007/JHEP02(2024)159",
    journal = "JHEP",
    volume = "02",
    pages = "159",
    year = "2024"
}

@article{Maldacena:1997re,
    author = "Maldacena, Juan Martin",
    title = "{The Large N limit of superconformal field theories and supergravity}",
    eprint = "hep-th/9711200",
    archivePrefix = "arXiv",
    reportNumber = "HUTP-97-A097, HUTP-98-A097",
    doi = "10.1023/A:1026654312961",
    journal = "Adv. Theor. Math. Phys.",
    volume = "2",
    pages = "231--252",
    year = "1998"
}

@article{Gubser:1998bc,
    author = "Gubser, S. S. and Klebanov, Igor R. and Polyakov, Alexander M.",
    title = "{Gauge theory correlators from noncritical string theory}",
    eprint = "hep-th/9802109",
    archivePrefix = "arXiv",
    reportNumber = "PUPT-1767",
    doi = "10.1016/S0370-2693(98)00377-3",
    journal = "Phys. Lett. B",
    volume = "428",
    pages = "105--114",
    year = "1998"
}

@article{Witten:1998qj,
    author = "Witten, Edward",
    title = "{Anti-de Sitter space and holography}",
    eprint = "hep-th/9802150",
    archivePrefix = "arXiv",
    reportNumber = "IASSNS-HEP-98-15",
    doi = "10.4310/ATMP.1998.v2.n2.a2",
    journal = "Adv. Theor. Math. Phys.",
    volume = "2",
    pages = "253--291",
    year = "1998"
}

@article{Aharony:1999ti,
    author = "Aharony, Ofer and Gubser, Steven S. and Maldacena, Juan Martin and Ooguri, Hirosi and Oz, Yaron",
    title = "{Large N field theories, string theory and gravity}",
    eprint = "hep-th/9905111",
    archivePrefix = "arXiv",
    reportNumber = "CERN-TH-99-122, HUTP-99-A027, LBNL-43113, RU-99-18, UCB-PTH-99-16, LBL-43113",
    doi = "10.1016/S0370-1573(99)00083-6",
    journal = "Phys. Rept.",
    volume = "323",
    pages = "183--386",
    year = "2000"
}

@article{Witten:1998zw,
    author = "Witten, Edward",
    editor = "Bergstrom, L. and Lindstrom, U.",
    title = "{Anti-de Sitter space, thermal phase transition, and confinement in gauge theories}",
    eprint = "hep-th/9803131",
    archivePrefix = "arXiv",
    reportNumber = "IASSNS-HEP-98-21",
    doi = "10.4310/ATMP.1998.v2.n3.a3",
    journal = "Adv. Theor. Math. Phys.",
    volume = "2",
    pages = "505--532",
    year = "1998"
}

@article{Klebanov:2000hb,
    author = "Klebanov, Igor R. and Strassler, Matthew J.",
    title = "{Supergravity and a confining gauge theory: Duality cascades and chi SB resolution of naked singularities}",
    eprint = "hep-th/0007191",
    archivePrefix = "arXiv",
    reportNumber = "IASSNS-HEP-00-56, PUPT-1944",
    doi = "10.1088/1126-6708/2000/08/052",
    journal = "JHEP",
    volume = "08",
    pages = "052",
    year = "2000"
}

@article{Maldacena:2000yy,
    author = "Maldacena, Juan Martin and Nunez, Carlos",
    title = "{Towards the large N limit of pure N=1 superYang-Mills}",
    eprint = "hep-th/0008001",
    archivePrefix = "arXiv",
    doi = "10.1103/PhysRevLett.86.588",
    journal = "Phys. Rev. Lett.",
    volume = "86",
    pages = "588--591",
    year = "2001"
}

@article{Babington:2003vm,
    author = "Babington, J. and Erdmenger, J. and Evans, Nick J. and Guralnik, Z. and Kirsch, I.",
    title = "{Chiral symmetry breaking and pions in nonsupersymmetric gauge / gravity duals}",
    eprint = "hep-th/0306018",
    archivePrefix = "arXiv",
    reportNumber = "HU-EP-03-27, SHEP-03-10",
    doi = "10.1103/PhysRevD.69.066007",
    journal = "Phys. Rev. D",
    volume = "69",
    pages = "066007",
    year = "2004"
}

@article{Chamseddine:1997nm,
    author = "Chamseddine, Ali H. and Volkov, Mikhail S.",
    title = "{NonAbelian BPS monopoles in N=4 gauged supergravity}",
    eprint = "hep-th/9707176",
    archivePrefix = "arXiv",
    reportNumber = "ETH-TH-97-16A, ZU-TH-97-15",
    doi = "10.1103/PhysRevLett.79.3343",
    journal = "Phys. Rev. Lett.",
    volume = "79",
    pages = "3343--3346",
    year = "1997"
}

@article{Butti:2004pk,
    author = "Butti, Agostino and Grana, Mariana and Minasian, Ruben and Petrini, Michela and Zaffaroni, Alberto",
    title = "{The Baryonic branch of Klebanov-Strassler solution: A supersymmetric family of SU(3) structure backgrounds}",
    eprint = "hep-th/0412187",
    archivePrefix = "arXiv",
    reportNumber = "BICOCCA-FT-04-18, CPHT-RR-070-1204, LPTENS-04-52",
    doi = "10.1088/1126-6708/2005/03/069",
    journal = "JHEP",
    volume = "03",
    pages = "069",
    year = "2005"
}

@article{Brower:2000rp,
    author = "Brower, Richard C. and Mathur, Samir D. and Tan, Chung-I",
    title = "{Glueball spectrum for QCD from AdS supergravity duality}",
    eprint = "hep-th/0003115",
    archivePrefix = "arXiv",
    reportNumber = "BROWN-HET-1217",
    doi = "10.1016/S0550-3213(00)00435-1",
    journal = "Nucl. Phys. B",
    volume = "587",
    pages = "249--276",
    year = "2000"
}

@article{Karch:2002sh,
    author = "Karch, Andreas and Katz, Emanuel",
    title = "{Adding flavor to AdS / CFT}",
    eprint = "hep-th/0205236",
    archivePrefix = "arXiv",
    reportNumber = "UW-PT-02-10",
    doi = "10.1088/1126-6708/2002/06/043",
    journal = "JHEP",
    volume = "06",
    pages = "043",
    year = "2002"
}

@article{Kruczenski:2003be,
    author = "Kruczenski, Martin and Mateos, David and Myers, Robert C. and Winters, David J.",
    title = "{Meson spectroscopy in AdS / CFT with flavor}",
    eprint = "hep-th/0304032",
    archivePrefix = "arXiv",
    doi = "10.1088/1126-6708/2003/07/049",
    journal = "JHEP",
    volume = "07",
    pages = "049",
    year = "2003"
}

@article{Sakai:2004cn,
    author = "Sakai, Tadakatsu and Sugimoto, Shigeki",
    title = "{Low energy hadron physics in holographic QCD}",
    eprint = "hep-th/0412141",
    archivePrefix = "arXiv",
    reportNumber = "IU-MSTP-63, YITP-04-70",
    doi = "10.1143/PTP.113.843",
    journal = "Prog. Theor. Phys.",
    volume = "113",
    pages = "843--882",
    year = "2005"
}

@article{Sakai:2005yt,
    author = "Sakai, Tadakatsu and Sugimoto, Shigeki",
    title = "{More on a holographic dual of QCD}",
    eprint = "hep-th/0507073",
    archivePrefix = "arXiv",
    reportNumber = "YITP-05-36, IU-MSTP-71",
    doi = "10.1143/PTP.114.1083",
    journal = "Prog. Theor. Phys.",
    volume = "114",
    pages = "1083--1118",
    year = "2005"
}

@article{Bigazzi:2020phm,
    author = "Bigazzi, Francesco and Caddeo, Alessio and Cotrone, Aldo L. and Paredes, Angel",
    title = "{Fate of false vacua in holographic first-order phase transitions}",
    eprint = "2008.02579",
    archivePrefix = "arXiv",
    primaryClass = "hep-th",
    doi = "10.1007/JHEP12(2020)200",
    journal = "JHEP",
    volume = "12",
    pages = "200",
    year = "2020"
}

@article{Ares:2020lbt,
    author = {Ares, F\"eanor Reuben and Hindmarsh, Mark and Hoyos, Carlos and Jokela, Niko},
    title = "{Gravitational waves from a holographic phase transition}",
    eprint = "2011.12878",
    archivePrefix = "arXiv",
    primaryClass = "hep-th",
    reportNumber = "HIP-2020-31/TH, Sussex-94886",
    doi = "10.1007/JHEP04(2021)100",
    journal = "JHEP",
    volume = "21",
    pages = "100",
    year = "2020"
}

@article{Bea:2021zsu,
    author = "Bea, Yago and Casalderrey-Solana, Jorge and Giannakopoulos, Thanasis and Mateos, David and Sanchez-Garitaonandia, Mikel and Zilh\~ao, Miguel",
    title = "{Bubble wall velocity from holography}",
    eprint = "2104.05708",
    archivePrefix = "arXiv",
    primaryClass = "hep-th",
    doi = "10.1103/PhysRevD.104.L121903",
    journal = "Phys. Rev. D",
    volume = "104",
    number = "12",
    pages = "L121903",
    year = "2021"
}

@article{Bigazzi:2021ucw,
    author = "Bigazzi, Francesco and Caddeo, Alessio and Canneti, Tommaso and Cotrone, Aldo L.",
    title = "{Bubble wall velocity at strong coupling}",
    eprint = "2104.12817",
    archivePrefix = "arXiv",
    primaryClass = "hep-ph",
    doi = "10.1007/JHEP08(2021)090",
    journal = "JHEP",
    volume = "08",
    pages = "090",
    year = "2021"
}

@article{Henriksson:2021zei,
    author = "Henriksson, Oscar",
    title = "{Black brane evaporation through D-brane bubble nucleation}",
    eprint = "2106.13254",
    archivePrefix = "arXiv",
    primaryClass = "hep-th",
    reportNumber = "HIP-2021-20/TH",
    doi = "10.1103/PhysRevD.105.L041901",
    journal = "Phys. Rev. D",
    volume = "105",
    number = "4",
    pages = "L041901",
    year = "2022"
}

@article{Ares:2021ntv,
    author = {Ares, F\"eanor Reuben and Henriksson, Oscar and Hindmarsh, Mark and Hoyos, Carlos and Jokela, Niko},
    title = "{Effective actions and bubble nucleation from holography}",
    eprint = "2109.13784",
    archivePrefix = "arXiv",
    primaryClass = "hep-th",
    reportNumber = "HIP-2021-30/TH",
    doi = "10.1103/PhysRevD.105.066020",
    journal = "Phys. Rev. D",
    volume = "105",
    number = "6",
    pages = "066020",
    year = "2022"
}

@article{Ares:2021nap,
    author = {Ares, F\"eanor Reuben and Henriksson, Oscar and Hindmarsh, Mark and Hoyos, Carlos and Jokela, Niko},
    title = "{Gravitational Waves at Strong Coupling from an Effective Action}",
    eprint = "2110.14442",
    archivePrefix = "arXiv",
    primaryClass = "hep-th",
    reportNumber = "HIP-2021-24/TH",
    doi = "10.1103/PhysRevLett.128.131101",
    journal = "Phys. Rev. Lett.",
    volume = "128",
    number = "13",
    pages = "131101",
    year = "2022"
}

@article{Morgante:2022zvc,
    author = "Morgante, Enrico and Ramberg, Nicklas and Schwaller, Pedro",
    title = "{Gravitational waves from dark SU(3) Yang-Mills theory}",
    eprint = "2210.11821",
    archivePrefix = "arXiv",
    primaryClass = "hep-ph",
    doi = "10.1103/PhysRevD.107.036010",
    journal = "Phys. Rev. D",
    volume = "107",
    number = "3",
    pages = "036010",
    year = "2023"
}

@article{Strassler:2006im,
    author = "Strassler, Matthew J. and Zurek, Kathryn M.",
    title = "{Echoes of a hidden valley at hadron colliders}",
    eprint = "hep-ph/0604261",
    archivePrefix = "arXiv",
    doi = "10.1016/j.physletb.2007.06.055",
    journal = "Phys. Lett. B",
    volume = "651",
    pages = "374--379",
    year = "2007"
}

@article{Cheung:2007ut,
    author = "Cheung, Kingman and Yuan, Tzu-Chiang",
    title = "{Hidden fermion as milli-charged dark matter in Stueckelberg Z- prime model}",
    eprint = "hep-ph/0701107",
    archivePrefix = "arXiv",
    doi = "10.1088/1126-6708/2007/03/120",
    journal = "JHEP",
    volume = "03",
    pages = "120",
    year = "2007"
}

@article{Hambye:2008bq,
    author = "Hambye, Thomas",
    title = "{Hidden vector dark matter}",
    eprint = "0811.0172",
    archivePrefix = "arXiv",
    primaryClass = "hep-ph",
    reportNumber = "ULB-TH-08-35",
    doi = "10.1088/1126-6708/2009/01/028",
    journal = "JHEP",
    volume = "01",
    pages = "028",
    year = "2009"
}

@article{Feng:2009mn,
    author = "Feng, Jonathan L. and Kaplinghat, Manoj and Tu, Huitzu and Yu, Hai-Bo",
    title = "{Hidden Charged Dark Matter}",
    eprint = "0905.3039",
    archivePrefix = "arXiv",
    primaryClass = "hep-ph",
    reportNumber = "UCI-TR-2009-06",
    doi = "10.1088/1475-7516/2009/07/004",
    journal = "JCAP",
    volume = "07",
    pages = "004",
    year = "2009"
}

@article{Cohen:2010kn,
    author = "Cohen, Timothy and Phalen, Daniel J. and Pierce, Aaron and Zurek, Kathryn M.",
    title = "{Asymmetric Dark Matter from a GeV Hidden Sector}",
    eprint = "1005.1655",
    archivePrefix = "arXiv",
    primaryClass = "hep-ph",
    reportNumber = "MCTP-10-18",
    doi = "10.1103/PhysRevD.82.056001",
    journal = "Phys. Rev. D",
    volume = "82",
    pages = "056001",
    year = "2010"
}

@article{Foot:2014uba,
    author = "Foot, R. and Vagnozzi, S.",
    title = "{Dissipative hidden sector dark matter}",
    eprint = "1409.7174",
    archivePrefix = "arXiv",
    primaryClass = "hep-ph",
    doi = "10.1103/PhysRevD.91.023512",
    journal = "Phys. Rev. D",
    volume = "91",
    pages = "023512",
    year = "2015"
}

@article{Bertone:2016nfn,
    author = "Bertone, Gianfranco and Hooper, Dan",
    title = "{History of dark matter}",
    eprint = "1605.04909",
    archivePrefix = "arXiv",
    primaryClass = "astro-ph.CO",
    reportNumber = "FERMILAB-PUB-16-157-A",
    doi = "10.1103/RevModPhys.90.045002",
    journal = "Rev. Mod. Phys.",
    volume = "90",
    number = "4",
    pages = "045002",
    year = "2018"
}

@article{DelNobile:2011je,
    author = "Del Nobile, Eugenio and Kouvaris, Chris and Sannino, Francesco",
    title = "{Interfering Composite Asymmetric Dark Matter for DAMA and CoGeNT}",
    eprint = "1105.5431",
    archivePrefix = "arXiv",
    primaryClass = "hep-ph",
    reportNumber = "CP3-ORIGINS-2011-18-\&-DIAS-2011-04",
    doi = "10.1103/PhysRevD.84.027301",
    journal = "Phys. Rev. D",
    volume = "84",
    pages = "027301",
    year = "2011"
}

@article{Hietanen:2013fya,
    author = "Hietanen, Ari and Lewis, Randy and Pica, Claudio and Sannino, Francesco",
    title = "{Composite Goldstone Dark Matter: Experimental Predictions from the Lattice}",
    eprint = "1308.4130",
    archivePrefix = "arXiv",
    primaryClass = "hep-ph",
    reportNumber = "CP3-ORIGINS-2013-30-DNRF90-\&-DIAS-2013-30",
    doi = "10.1007/JHEP12(2014)130",
    journal = "JHEP",
    volume = "12",
    pages = "130",
    year = "2014"
}

@article{Cline:2016nab,
    author = "Cline, James M. and Huang, Weicong and Moore, Guy D.",
    title = "{Challenges for models with composite states}",
    eprint = "1607.07865",
    archivePrefix = "arXiv",
    primaryClass = "hep-ph",
    doi = "10.1103/PhysRevD.94.055029",
    journal = "Phys. Rev. D",
    volume = "94",
    number = "5",
    pages = "055029",
    year = "2016"
}

@article{Cacciapaglia:2020kgq,
    author = "Cacciapaglia, Giacomo and Pica, Claudio and Sannino, Francesco",
    title = "{Fundamental Composite Dynamics: A Review}",
    eprint = "2002.04914",
    archivePrefix = "arXiv",
    primaryClass = "hep-ph",
    doi = "10.1016/j.physrep.2020.07.002",
    journal = "Phys. Rept.",
    volume = "877",
    pages = "1--70",
    year = "2020"
}

@article{Dondi:2019olm,
    author = "Dondi, Nicola Andrea and Sannino, Francesco and Smirnov, Juri",
    title = "{Thermal history of composite dark matter}",
    eprint = "1905.08810",
    archivePrefix = "arXiv",
    primaryClass = "hep-ph",
    doi = "10.1103/PhysRevD.101.103010",
    journal = "Phys. Rev. D",
    volume = "101",
    number = "10",
    pages = "103010",
    year = "2020"
}

@article{Ge:2019voa,
    author = "Ge, Shuailiang and Lawson, Kyle and Zhitnitsky, Ariel",
    title = "{Axion quark nugget dark matter model: Size distribution and survival pattern}",
    eprint = "1903.05090",
    archivePrefix = "arXiv",
    primaryClass = "hep-ph",
    doi = "10.1103/PhysRevD.99.116017",
    journal = "Phys. Rev. D",
    volume = "99",
    number = "11",
    pages = "116017",
    year = "2019"
}

@article{Beylin:2019gtw,
    author = "Beylin, Vitaly and Khlopov, Maxim Yu. and Kuksa, Vladimir and Volchanskiy, Nikolay",
    title = "{Hadronic and Hadron-Like Physics of Dark Matter}",
    eprint = "1904.12013",
    archivePrefix = "arXiv",
    primaryClass = "hep-ph",
    doi = "10.3390/sym11040587",
    journal = "Symmetry",
    volume = "11",
    number = "4",
    pages = "587",
    year = "2019"
}

@article{Yamanaka:2019aeq,
    author = "Yamanaka, Nodoka and Iida, Hideaki and Nakamura, Atsushi and Wakayama, Masayuki",
    title = "{Dark matter scattering cross section and dynamics in dark Yang-Mills theory}",
    eprint = "1910.01440",
    archivePrefix = "arXiv",
    primaryClass = "hep-ph",
    doi = "10.1016/j.physletb.2020.136056",
    journal = "Phys. Lett. B",
    volume = "813",
    pages = "136056",
    year = "2021"
}

@article{Yamanaka:2019yek,
    author = "Yamanaka, Nodoka and Iida, Hideaki and Nakamura, Atsushi and Wakayama, Masayuki",
    title = "{Glueball scattering cross section in lattice SU(2) Yang-Mills theory}",
    eprint = "1910.07756",
    archivePrefix = "arXiv",
    primaryClass = "hep-lat",
    doi = "10.1103/PhysRevD.102.054507",
    journal = "Phys. Rev. D",
    volume = "102",
    number = "5",
    pages = "054507",
    year = "2020"
}

@article{Cai:2020njb,
    author = "Cai, Haiying and Cacciapaglia, Giacomo",
    title = "{Singlet dark matter in the SU(6)/SO(6) composite Higgs model}",
    eprint = "2007.04338",
    archivePrefix = "arXiv",
    primaryClass = "hep-ph",
    reportNumber = "APCTP Pre2020-015, APCTP Pre2020-015",
    doi = "10.1103/PhysRevD.103.055002",
    journal = "Phys. Rev. D",
    volume = "103",
    number = "5",
    pages = "055002",
    year = "2021"
}

@article{Hochberg:2014dra,
    author = "Hochberg, Yonit and Kuflik, Eric and Volansky, Tomer and Wacker, Jay G.",
    title = "{Mechanism for Thermal Relic Dark Matter of Strongly Interacting Massive Particles}",
    eprint = "1402.5143",
    archivePrefix = "arXiv",
    primaryClass = "hep-ph",
    doi = "10.1103/PhysRevLett.113.171301",
    journal = "Phys. Rev. Lett.",
    volume = "113",
    pages = "171301",
    year = "2014"
}

@article{Hochberg:2014kqa,
    author = "Hochberg, Yonit and Kuflik, Eric and Murayama, Hitoshi and Volansky, Tomer and Wacker, Jay G.",
    title = "{Model for Thermal Relic Dark Matter of Strongly Interacting Massive Particles}",
    eprint = "1411.3727",
    archivePrefix = "arXiv",
    primaryClass = "hep-ph",
    doi = "10.1103/PhysRevLett.115.021301",
    journal = "Phys. Rev. Lett.",
    volume = "115",
    number = "2",
    pages = "021301",
    year = "2015"
}

@article{Hochberg:2015vrg,
    author = "Hochberg, Yonit and Kuflik, Eric and Murayama, Hitoshi",
    title = "{SIMP Spectroscopy}",
    eprint = "1512.07917",
    archivePrefix = "arXiv",
    primaryClass = "hep-ph",
    reportNumber = "UCB-PTH-15-16, IPMU15-0218",
    doi = "10.1007/JHEP05(2016)090",
    journal = "JHEP",
    volume = "05",
    pages = "090",
    year = "2016"
}

@article{Bernal:2017mqb,
    author = "Bernal, Nicol\'as and Chu, Xiaoyong and Pradler, Josef",
    title = "{Simply split strongly interacting massive particles}",
    eprint = "1702.04906",
    archivePrefix = "arXiv",
    primaryClass = "hep-ph",
    doi = "10.1103/PhysRevD.95.115023",
    journal = "Phys. Rev. D",
    volume = "95",
    number = "11",
    pages = "115023",
    year = "2017"
}

@article{Berlin:2018tvf,
    author = "Berlin, Asher and Blinov, Nikita and Gori, Stefania and Schuster, Philip and Toro, Natalia",
    title = "{Cosmology and Accelerator Tests of Strongly Interacting Dark Matter}",
    eprint = "1801.05805",
    archivePrefix = "arXiv",
    primaryClass = "hep-ph",
    reportNumber = "SLAC-PUB-17135",
    doi = "10.1103/PhysRevD.97.055033",
    journal = "Phys. Rev. D",
    volume = "97",
    number = "5",
    pages = "055033",
    year = "2018"
}

@article{Bernal:2019uqr,
    author = "Bernal, Nicol\'as and Chu, Xiaoyong and Kulkarni, Suchita and Pradler, Josef",
    title = "{Self-interacting dark matter without prejudice}",
    eprint = "1912.06681",
    archivePrefix = "arXiv",
    primaryClass = "hep-ph",
    reportNumber = "PI/UAN-2019-663FT",
    doi = "10.1103/PhysRevD.101.055044",
    journal = "Phys. Rev. D",
    volume = "101",
    number = "5",
    pages = "055044",
    year = "2020"
}

@article{Tsai:2020vpi,
    author = "Tsai, Yu-Dai and McGehee, Robert and Murayama, Hitoshi",
    title = "{Resonant Self-Interacting Dark Matter from Dark QCD}",
    eprint = "2008.08608",
    archivePrefix = "arXiv",
    primaryClass = "hep-ph",
    reportNumber = "FERMILAB-PUB-20-365-AE-T",
    doi = "10.1103/PhysRevLett.128.172001",
    journal = "Phys. Rev. Lett.",
    volume = "128",
    number = "17",
    pages = "172001",
    year = "2022"
}

@article{Kondo:2022lgg,
    author = "Kondo, Dan and McGehee, Robert and Melia, Tom and Murayama, Hitoshi",
    title = "{Linear sigma dark matter}",
    eprint = "2205.08088",
    archivePrefix = "arXiv",
    primaryClass = "hep-ph",
    reportNumber = "LCTP-22-06",
    doi = "10.1007/JHEP09(2022)041",
    journal = "JHEP",
    volume = "09",
    pages = "041",
    year = "2022"
}

@article{Chu:2024rrv,
    author = "Chu, Xiaoyong and Nikolic, Marco and Pradler, Josef",
    title = "{Even SIMP miracles are possible}",
    eprint = "2401.12283",
    archivePrefix = "arXiv",
    primaryClass = "hep-ph",
    doi = "10.1103/PhysRevLett.133.021003",
    journal = "Phys. Rev. Lett.",
    volume = "133",
    number = "2",
    pages = "2",
    year = "2024"
}

@article{Appelquist:2015yfa,
    author = "Appelquist, Thomas and others",
    title = "{Stealth Dark Matter: Dark scalar baryons through the Higgs portal}",
    eprint = "1503.04203",
    archivePrefix = "arXiv",
    primaryClass = "hep-ph",
    reportNumber = "INT-PUB-15-004, LLNL-JRNL-667446",
    doi = "10.1103/PhysRevD.92.075030",
    journal = "Phys. Rev. D",
    volume = "92",
    number = "7",
    pages = "075030",
    year = "2015"
}

@article{Appelquist:2015zfa,
    author = "Appelquist, Thomas and others",
    title = "{Detecting Stealth Dark Matter Directly through Electromagnetic Polarizability}",
    eprint = "1503.04205",
    archivePrefix = "arXiv",
    primaryClass = "hep-ph",
    reportNumber = "INT-PUB-15-005, LLNL-JRNL-667121",
    doi = "10.1103/PhysRevLett.115.171803",
    journal = "Phys. Rev. Lett.",
    volume = "115",
    number = "17",
    pages = "171803",
    year = "2015"
}

@article{LatticeStrongDynamics:2020jwi,
    author = "Brower, R. C. and others",
    collaboration = "Lattice Strong Dynamics",
    title = "{Stealth dark matter confinement transition and gravitational waves}",
    eprint = "2006.16429",
    archivePrefix = "arXiv",
    primaryClass = "hep-lat",
    reportNumber = "LLNL-JRNL-811356; RIKEN-iTHEMS-Report-20",
    doi = "10.1103/PhysRevD.103.014505",
    journal = "Phys. Rev. D",
    volume = "103",
    number = "1",
    pages = "014505",
    year = "2021"
}

@article{Svetitsky:1982gs,
    author = "Svetitsky, Benjamin and Yaffe, Laurence G.",
    title = "{Critical Behavior at Finite Temperature Confinement Transitions}",
    reportNumber = "CLNS-82/530, NSF-ITP-82-53",
    doi = "10.1016/0550-3213(82)90172-9",
    journal = "Nucl. Phys. B",
    volume = "210",
    pages = "423--447",
    year = "1982"
}

@article{Yaffe:1982qf,
    author = "Yaffe, L. G. and Svetitsky, B.",
    title = "{First Order Phase Transition in the SU(3) Gauge Theory at Finite Temperature}",
    reportNumber = "CALT-68-913",
    doi = "10.1103/PhysRevD.26.963",
    journal = "Phys. Rev. D",
    volume = "26",
    pages = "963",
    year = "1982"
}

@article{Saito:2011fs,
    author = "Saito, H. and Ejiri, S. and Aoki, S. and Hatsuda, T. and Kanaya, K. and Maezawa, Y. and Ohno, H. and Umeda, T.",
    collaboration = "WHOT-QCD",
    title = "{Phase structure of finite temperature QCD in the heavy quark region}",
    eprint = "1106.0974",
    archivePrefix = "arXiv",
    primaryClass = "hep-lat",
    reportNumber = "UTHEP-629, RIKEN-MP-22",
    doi = "10.1103/PhysRevD.85.079902",
    journal = "Phys. Rev. D",
    volume = "84",
    pages = "054502",
    year = "2011",
    note = "[Erratum: Phys.Rev.D 85, 079902 (2012)]"
}

@article{Ejiri:2019csa,
    author = "Ejiri, Shinji and Itagaki, Shota and Iwami, Ryo and Kanaya, Kazuyuki and Kitazawa, Masakiyo and Kiyohara, Atsushi and Shirogane, Mizuki and Umeda, Takashi",
    collaboration = "WHOT-QCD",
    title = "{End point of the first-order phase transition of QCD in the heavy quark region by reweighting from quenched QCD}",
    eprint = "1912.10500",
    archivePrefix = "arXiv",
    primaryClass = "hep-lat",
    reportNumber = "UTHEP-742, J-PARC-TH-0209",
    doi = "10.1103/PhysRevD.101.054505",
    journal = "Phys. Rev. D",
    volume = "101",
    number = "5",
    pages = "054505",
    year = "2020"
}

@article{Kiyohara:2021smr,
    author = "Kiyohara, Atsushi and Kitazawa, Masakiyo and Ejiri, Shinji and Kanaya, Kazuyuki",
    title = "{Finite-size scaling around the critical point in the heavy quark region of QCD}",
    eprint = "2108.00118",
    archivePrefix = "arXiv",
    primaryClass = "hep-lat",
    reportNumber = "J-PARC-TH-0246,UTHEP-758",
    doi = "10.1103/PhysRevD.104.114509",
    journal = "Phys. Rev. D",
    volume = "104",
    number = "11",
    pages = "114509",
    year = "2021"
}

@article{Fromm:2011qi,
    author = "Fromm, Michael and Langelage, Jens and Lottini, Stefano and Philipsen, Owe",
    title = "{The QCD deconfinement transition for heavy quarks and all baryon chemical potentials}",
    eprint = "1111.4953",
    archivePrefix = "arXiv",
    primaryClass = "hep-lat",
    doi = "10.1007/JHEP01(2012)042",
    journal = "JHEP",
    volume = "01",
    pages = "042",
    year = "2012"
}

@article{Cuteri:2020yke,
    author = {Cuteri, Francesca and Philipsen, Owe and Sch\"on, Alena and Sciarra, Alessandro},
    title = "{Deconfinement critical point of lattice QCD with $N_f$=2 Wilson fermions}",
    eprint = "2009.14033",
    archivePrefix = "arXiv",
    primaryClass = "hep-lat",
    doi = "10.1103/PhysRevD.103.014513",
    journal = "Phys. Rev. D",
    volume = "103",
    number = "1",
    pages = "014513",
    year = "2021"
}

@article{Borsanyi:2021yoz,
    author = "Borsanyi, Szabolcs and Fodor, Zoltan and Guenther, Jana N. and Kara, Ruben and Parotto, Paolo and Pasztor, Attila and Sexty, Denes",
    title = "{The upper right corner of the Columbia plot with staggered fermions}",
    eprint = "2112.04192",
    archivePrefix = "arXiv",
    primaryClass = "hep-lat",
    doi = "10.22323/1.396.0496",
    journal = "PoS",
    volume = "LATTICE2021",
    pages = "496",
    year = "2022"
}

@inproceedings{Aarts:2023vsf,
    author = "Aarts, Gert and others",
    title = "{Phase Transitions in Particle Physics - Results and Perspectives from Lattice Quantum Chromo-Dynamics}",
    booktitle = "{Phase Transitions in Particle Physics}: {Results and Perspectives from Lattice Quantum Chromo-Dynamics}",
    eprint = "2301.04382",
    archivePrefix = "arXiv",
    primaryClass = "hep-lat",
    month = "1",
    year = "2023"
}

@article{Kajantie:1981wh,
    author = "Kajantie, K. and Montonen, C. and Pietarinen, E.",
    title = "{Phase Transition of SU(3) Gauge Theory at Finite Temperature}",
    reportNumber = "HU-TFT-81-8",
    doi = "10.1007/BF01410665",
    journal = "Z. Phys. C",
    volume = "9",
    pages = "253",
    year = "1981"
}

@article{Celik:1983wz,
    author = "Celik, T. and Engels, J. and Satz, H.",
    title = "{The Order of the Deconfinement Transition in SU(3) Yang-Mills Theory}",
    reportNumber = "BI-TP-83-04",
    doi = "10.1016/0370-2693(83)91314-X",
    journal = "Phys. Lett. B",
    volume = "125",
    pages = "411--414",
    year = "1983"
}

@article{Kogut:1983mn,
    author = "Kogut, John B. and Matsuoka, H. and Stone, M. and Wyld, H. W. and Shenker, Stephen H. and Shigemitsu, J. and Sinclair, D. K.",
    title = "{Quark and Gluon Latent Heats at the Deconfinement Phase Transition in SU(3) Gauge Theory}",
    reportNumber = "ILL-TH-83-9, DOE/ER/01545-333",
    doi = "10.1103/PhysRevLett.51.869",
    journal = "Phys. Rev. Lett.",
    volume = "51",
    pages = "869",
    year = "1983"
}

@article{Svetitsky:1983bq,
    author = "Svetitsky, Benjamin and Fucito, Francesco",
    title = "{Latent Heat of the SU(3) Gauge Theory}",
    reportNumber = "CLNS-83-571",
    doi = "10.1016/0370-2693(83)91112-7",
    journal = "Phys. Lett. B",
    volume = "131",
    pages = "165",
    year = "1983"
}

@article{Gottlieb:1985ug,
    author = "Gottlieb, Steven A. and Kuti, J. and Toussaint, D. and Kennedy, A. D. and Meyer, S. and Pendleton, B. J. and Sugar, R. L.",
    title = "{The Deconfining Phase Transition and the Continuum Limit of Lattice Quantum Chromodynamics}",
    reportNumber = "UCSD-10P10-250",
    doi = "10.1103/PhysRevLett.55.1958",
    journal = "Phys. Rev. Lett.",
    volume = "55",
    pages = "1958",
    year = "1985"
}

@article{Brown:1988qe,
    author = "Brown, F. R. and Christ, N. H. and Deng, Y. F. and Gao, M. S. and Woch, T. J.",
    title = "{Nature of the Deconfining Phase Transition in SU(3) Lattice Gauge Theory}",
    doi = "10.1103/PhysRevLett.61.2058",
    journal = "Phys. Rev. Lett.",
    volume = "61",
    pages = "2058",
    year = "1988"
}

@article{Fukugita:1989yb,
    author = "Fukugita, M. and Okawa, M. and Ukawa, A.",
    title = "{Order of the Deconfining Phase Transition in SU(3) Lattice Gauge Theory}",
    reportNumber = "KEK-TH-228, KEK-Preprint-89-28",
    doi = "10.1103/PhysRevLett.63.1768",
    journal = "Phys. Rev. Lett.",
    volume = "63",
    pages = "1768",
    year = "1989"
}

@article{Bacilieri:1989ir,
    author = "Bacilieri, P. and others",
    title = "{A New Computation of the Correlation Length Near the Deconfining Transition in SU(3)}",
    doi = "10.1016/0370-2693(89)91241-0",
    journal = "Phys. Lett. B",
    volume = "224",
    pages = "333--338",
    year = "1989"
}

@article{Alves:1990pn,
    author = "Alves, Nelson A. and Berg, Bernd A. and Sanielevici, Sergiu",
    title = "{Binder Energy Cumulant for SU(3) Lattice Gauge Theory}",
    reportNumber = "FSU-SCRI-90-11",
    doi = "10.1016/0370-2693(90)91869-D",
    journal = "Phys. Lett. B",
    volume = "241",
    pages = "557--560",
    year = "1990"
}

@article{Boyd:1995zg,
    author = "Boyd, G. and Engels, J. and Karsch, F. and Laermann, E. and Legeland, C. and Lutgemeier, M. and Petersson, B.",
    title = "{Equation of state for the SU(3) gauge theory}",
    eprint = "hep-lat/9506025",
    archivePrefix = "arXiv",
    reportNumber = "BI-TP-95-23",
    doi = "10.1103/PhysRevLett.75.4169",
    journal = "Phys. Rev. Lett.",
    volume = "75",
    pages = "4169--4172",
    year = "1995"
}

@article{Boyd:1996bx,
    author = "Boyd, G. and Engels, J. and Karsch, F. and Laermann, E. and Legeland, C. and Lutgemeier, M. and Petersson, B.",
    title = "{Thermodynamics of SU(3) lattice gauge theory}",
    eprint = "hep-lat/9602007",
    archivePrefix = "arXiv",
    reportNumber = "BI-TP-96-04",
    doi = "10.1016/0550-3213(96)00170-8",
    journal = "Nucl. Phys. B",
    volume = "469",
    pages = "419--444",
    year = "1996"
}

@article{Borsanyi:2012ve,
    author = "Borsanyi, Sz. and Endrodi, G. and Fodor, Z. and Katz, S. D. and Szabo, K. K.",
    title = "{Precision SU(3) lattice thermodynamics for a large temperature range}",
    eprint = "1204.6184",
    archivePrefix = "arXiv",
    primaryClass = "hep-lat",
    reportNumber = "WUB-12-09",
    doi = "10.1007/JHEP07(2012)056",
    journal = "JHEP",
    volume = "07",
    pages = "056",
    year = "2012"
}

@article{Shirogane:2016zbf,
    author = "Shirogane, Mizuki and Ejiri, Shinji and Iwami, Ryo and Kanaya, Kazuyuki and Kitazawa, Masakiyo",
    title = "{Latent heat at the first order phase transition point of SU(3) gauge theory}",
    eprint = "1605.02997",
    archivePrefix = "arXiv",
    primaryClass = "hep-lat",
    doi = "10.1103/PhysRevD.94.014506",
    journal = "Phys. Rev. D",
    volume = "94",
    number = "1",
    pages = "014506",
    year = "2016"
}

@article{Borsanyi:2022xml,
    author = "Borsanyi, S. and R., Kara and Fodor, Z. and Godzieba, D. A. and Parotto, P. and Sexty, D.",
    title = "{Precision study of the continuum SU(3) Yang-Mills theory: How to use parallel tempering to improve on supercritical slowing down for first order phase transitions}",
    eprint = "2202.05234",
    archivePrefix = "arXiv",
    primaryClass = "hep-lat",
    doi = "10.1103/PhysRevD.105.074513",
    journal = "Phys. Rev. D",
    volume = "105",
    number = "7",
    pages = "074513",
    year = "2022"
}

@article{Lucini:2002ku,
    author = "Lucini, Biagio and Teper, Michael and Wenger, Urs",
    title = "{The Deconfinement transition in SU(N) gauge theories}",
    eprint = "hep-lat/0206029",
    archivePrefix = "arXiv",
    reportNumber = "OUTP-02-28P",
    doi = "10.1016/S0370-2693(02)02556-X",
    journal = "Phys. Lett. B",
    volume = "545",
    pages = "197--206",
    year = "2002"
}

@article{Lucini:2003zr,
    author = "Lucini, Biagio and Teper, Michael and Wenger, Urs",
    title = "{The High temperature phase transition in SU(N) gauge theories}",
    eprint = "hep-lat/0307017",
    archivePrefix = "arXiv",
    reportNumber = "OUTP-03-19P",
    doi = "10.1088/1126-6708/2004/01/061",
    journal = "JHEP",
    volume = "01",
    pages = "061",
    year = "2004"
}

@article{Lucini:2005vg,
    author = "Lucini, Biagio and Teper, Michael and Wenger, Urs",
    title = "{Properties of the deconfining phase transition in SU(N) gauge theories}",
    eprint = "hep-lat/0502003",
    archivePrefix = "arXiv",
    reportNumber = "DESY-05-021",
    doi = "10.1088/1126-6708/2005/02/033",
    journal = "JHEP",
    volume = "02",
    pages = "033",
    year = "2005"
}

@article{Panero:2009tv,
    author = "Panero, Marco",
    title = "{Thermodynamics of the QCD plasma and the large-N limit}",
    eprint = "0907.3719",
    archivePrefix = "arXiv",
    primaryClass = "hep-lat",
    doi = "10.1103/PhysRevLett.103.232001",
    journal = "Phys. Rev. Lett.",
    volume = "103",
    pages = "232001",
    year = "2009"
}

@article{Datta:2010sq,
    author = "Datta, Saumen and Gupta, Sourendu",
    title = "{Continuum Thermodynamics of the Gluo$N_c$ Plasma}",
    eprint = "1006.0938",
    archivePrefix = "arXiv",
    primaryClass = "hep-lat",
    reportNumber = "TIFR-TH-10-13",
    doi = "10.1103/PhysRevD.82.114505",
    journal = "Phys. Rev. D",
    volume = "82",
    pages = "114505",
    year = "2010"
}

@article{Lucini:2012wq,
    author = "Lucini, Biagio and Rago, Antonio and Rinaldi, Enrico",
    title = "{SU($N_c$) gauge theories at deconfinement}",
    eprint = "1202.6684",
    archivePrefix = "arXiv",
    primaryClass = "hep-lat",
    doi = "10.1016/j.physletb.2012.04.070",
    journal = "Phys. Lett. B",
    volume = "712",
    pages = "279--283",
    year = "2012"
}

@article{Holland:2003kg,
    author = "Holland, K. and Pepe, M. and Wiese, U. J.",
    title = "{The Deconfinement phase transition of Sp(2) and Sp(3) Yang-Mills theories in (2+1)-dimensions and (3+1)-dimensions}",
    eprint = "hep-lat/0312022",
    archivePrefix = "arXiv",
    doi = "10.1016/j.nuclphysb.2004.06.026",
    journal = "Nucl. Phys. B",
    volume = "694",
    pages = "35--58",
    year = "2004"
}

@article{Pepe:2005sz,
    author = "Pepe, Michele",
    editor = "Alexandrou, Constantia and Panagopoulos, Haralambos and Schierholz, Gerrit",
    title = "{Confinement and the center of the gauge group}",
    eprint = "hep-lat/0510013",
    archivePrefix = "arXiv",
    doi = "10.1016/j.nuclphysbps.2006.01.045",
    journal = "PoS",
    volume = "LAT2005",
    pages = "017",
    year = "2006"
}

@article{Pepe:2006er,
    author = "Pepe, M. and Wiese, U. -J.",
    title = "{Exceptional Deconfinement in G(2) Gauge Theory}",
    eprint = "hep-lat/0610076",
    archivePrefix = "arXiv",
    doi = "10.1016/j.nuclphysb.2006.12.024",
    journal = "Nucl. Phys. B",
    volume = "768",
    pages = "21--37",
    year = "2007"
}

@article{Cossu:2007dk,
    author = "Cossu, Guido and D'Elia, Massimo and Di Giacomo, Adriano and Lucini, Biagio and Pica, Claudio",
    title = "{G(2) gauge theory at finite temperature}",
    eprint = "0709.0669",
    archivePrefix = "arXiv",
    primaryClass = "hep-lat",
    reportNumber = "IFUP-TH-2007-22, BNL-NT-07-36",
    doi = "10.1088/1126-6708/2007/10/100",
    journal = "JHEP",
    volume = "10",
    pages = "100",
    year = "2007"
}

@article{Bruno:2014rxa,
    author = "Bruno, Mattia and Caselle, Michele and Panero, Marco and Pellegrini, Roberto",
    title = "{Exceptional thermodynamics: the equation of state of G$_{2}$ gauge theory}",
    eprint = "1409.8305",
    archivePrefix = "arXiv",
    primaryClass = "hep-lat",
    reportNumber = "DESY-14-146, IFT-UAM-CSIC-14-076",
    doi = "10.1007/JHEP03(2015)057",
    journal = "JHEP",
    volume = "03",
    pages = "057",
    year = "2015"
}

@article{Pisarski:2000eq,
    author = "Pisarski, Robert D.",
    title = "{Quark gluon plasma as a condensate of SU(3) Wilson lines}",
    eprint = "hep-ph/0006205",
    archivePrefix = "arXiv",
    doi = "10.1103/PhysRevD.62.111501",
    journal = "Phys. Rev. D",
    volume = "62",
    pages = "111501",
    year = "2000"
}

@article{Pisarski:2001pe,
    author = "Pisarski, Robert D.",
    editor = "Karsch, F. and Satz, H.",
    title = "{Tests of the Polyakov loops model}",
    eprint = "hep-ph/0112037",
    archivePrefix = "arXiv",
    doi = "10.1016/S0375-9474(02)00699-1",
    journal = "Nucl. Phys. A",
    volume = "702",
    pages = "151--158",
    year = "2002"
}

@inproceedings{Pisarski:2002ji,
    author = "Pisarski, Robert D.",
    title = "{Notes on the deconfining phase transition}",
    booktitle = "{Cargese Summer School on QCD Perspectives on Hot and Dense Matter}",
    eprint = "hep-ph/0203271",
    archivePrefix = "arXiv",
    pages = "353--384",
    month = "3",
    year = "2002"
}

@article{Sannino:2002wb,
    author = "Sannino, Francesco",
    title = "{Polyakov loops versus hadronic states}",
    eprint = "hep-ph/0204174",
    archivePrefix = "arXiv",
    doi = "10.1103/PhysRevD.66.034013",
    journal = "Phys. Rev. D",
    volume = "66",
    pages = "034013",
    year = "2002"
}

@misc{Kondo:2015noa,
    author = "Kondo, Kei-Ichi",
    title = "{Confinement--deconfinement phase transition and gauge-invariant gluonic mass in Yang-Mills theory}",
    eprint = "1508.02656",
    archivePrefix = "arXiv",
    primaryClass = "hep-th",
    reportNumber = "CHIBA-EP-211, CHIBA-EP-211-V2",
    month = "8",
    year = "2015"
}

\end{document}